\documentclass[a4paper,12pt]{article}
\usepackage[english]{babel}
\usepackage{fancyhdr}
\usepackage[T1]{fontenc}
\usepackage{hyperref}
\usepackage{dsfont}
\usepackage{amsfonts}
\usepackage{amsmath}
\usepackage{amssymb}
\usepackage{amsthm}
\usepackage{eucal}
\usepackage{mathrsfs}
\usepackage[all]{xy}
\usepackage{units}
\usepackage[dvips]{graphicx}
\usepackage{color}
\numberwithin{equation}{section}

\usepackage{rotating}

\newtheorem{teo}{\textsc{Theorem}}[section]
\newtheorem{lem}[teo]{\textsc{Lemma}}
\newtheorem{propos}[teo]{\textsc{Proposition}}

\newtheorem*{teo*}{\textsc{Theorem}}
\newtheorem*{defi*}{\textsc{Definition}}
\newtheorem*{corol*}{\textsc{Corollary}}
\newtheorem*{conv*}{\textsc{Convention}}

\theoremstyle{definition}
\newtheorem*{ex*}{\textsc{Example}}
\newtheorem{rk}[teo]{\textsc{Remark}}
\newtheorem*{rk*}{\textsc{Remark}}
\newtheorem*{qst*}{\textsc{Question}}

\newcommand{\ii}{\,{\rm i}\,}
\newcommand{\Proof}{\begin{proof}[\textsc{\bf{Proof}}]}
\newcommand{\CVD}{\end{proof}}

\newcommand{\R}{\mathbb R}
\newcommand{\N}{\mathbb N}
\newcommand{\C}{\mathbb C}                           
\newcommand{\Q}{\mathbb Q}
\newcommand{\Z}{\mathbb Z}

\newcommand{\sss}[1]{\CMcal{#1}}
\newcommand{\bbb}[1]{\mathscr{#1}}
\newcommand{\rrr}[1]{\mathfrak{#1}}
\newcommand{\num}[1]{\mathds {#1}}

\newcommand{\bra}[1]{\langle #1|}
\newcommand{\ket}[1]{|#1\rangle}
\newcommand{\braket}[2]{\langle #1|#2\rangle}        
\newcommand{\ketbra}[2]{|#1\rangle\langle#2|}

\newcommand{\expo}[1]{\mbox{{\upshape e}}^{#1}}                 

\newcommand{\ncint}{\mathrel{{\ooalign{$\int$\cr\kern+.07em\raise.15ex\hbox{$\pmb{\scriptstyle-}$}\cr}}}\!\!}           
\newcommand{\nint}{\int\mkern-19mu-\;}

\newcommand{\ncpartial}{\mathrel{{\ooalign{$\partial$\cr\kern+.29em\raise.79ex\hbox{$\pmb{\scriptstyle-}$}\cr}}}\!\!}
\newcommand{\ncC}{\mathrel{{\ooalign{$C_1$\cr\kern-.13em\raise.45ex\hbox{$\pmb{\scriptstyle-}$}\cr}}}\!\!}

\newcommand{\virg}[1]{\lq\lq#1\rq\rq}                

\def\dd{{\rm d}}

\vspace{-5mm}
\title{\textbf{\Huge{Generalized TKNN-equations}}}
\author{\large{$\text{Giuseppe De Nittis}^*$ and $\text{Giovanni Landi}^{**}$}
\\
\\
\\
\normalsize{$^\ast$ LAGA, Institut Galil\'{e}e, Universit\'{e} Paris 13}\\
\normalsize{99, avenue J.-B. Cl\'{e}ment, F-93430 Villetaneuse, France.}\\
\footnotesize{\texttt{denittis@math.univ-paris13.fr}}\vspace{1mm}
\\
\\
\normalsize{$^{**}$ Dipartimento di Matematica e Informatica, Universit\`{a} di Trieste} \\
\normalsize{Via A. Valerio 12/1,
I-34127 Trieste, Italy}\\
\normalsize{and INFN, Sezione di Trieste, Trieste, Italy}\\
\footnotesize{\texttt{landi@units.it}}\\
}
\date{ 16 June 2011 }
\begin{document}

\maketitle

\begin{abstract}
We derive generalized TKNN-equations 
via bundle representations of the noncommutative torus with rational deformation parameter, the bundle coming from spectral projections in the torus algebra.  
These equations relate Chern numbers of dual bundles which we interpret as Hall conductances for Dirac-like Hamiltonians describing magnetic Bloch electrons in a strong magnetic field. We also present  their generalizations for irrational values of the deformation parameter.

\end{abstract}

\vfill

\newpage

\tableofcontents 

\parskip 1ex

\linespread{1.1}

\section{Introduction: TKNN-equations}\label{se:intro}

The Hall conductances associated to the energy spectrum of the single particle Hamiltonian operator  with periodic potential and magnetic field in the limit of a strong and a weak magnetic field are related by the TKNN-equations obtained for the first time in the seminal paper \cite{Thouless-Kohmoto-Nightingale-Nijs-82}.
Since their first derivation the common lore was to think of these conductances (in suitable units) as Chern numbers of vector bundles associated with the energy bands  of the Hamiltonian operator. However, this association has been rigorously established only recently \cite{denittis-10, denittis-faure-panati} with a systematic derivation of the bundles and the corresponding bundle representations coming from the symmetries of the physical system. 
The bundles are associated, via canonical representations  to spectral projections in the algebra of the noncommutative torus with rational deformation parameter.
In the present work these results are extended to general $(q,r)$-Weyl representations of the the noncommutative torus. The Chern numbers of the dual bundles are related by \emph{generalized TKNN-equations} which we interpret as relating Hall conductances of more general Dirac-like Hamiltonians for magnetic Bloch electrons. 

The physics of the \emph{integer quantum Hall effect} (IQHE)  reveals a variety of 
surprising and attractive features (\cite{morandi-88,Bellissard-Baldes-Elst-94,avron-04,graf-07} and references therein). The appearance of fractal spectra, quantization of the transverse conductance, anomalous thermodynamic phase diagrams and so on, are consequences of a subtle interplay between the crystal length scale and the magnetic length scale. In the last decades these phenomena have been the subject 
of several studies devoted to analytic and geometric properties of effective models. 

Indeed, the Schr\"{o}dinger operator for a single particle moving in a plane in a periodic potential and 
subject to an uniform orthogonal magnetic field of strength $B$ (\emph{magnetic Bloch electron}) is given by
\begin{equation}\label{eq_000}
H_B:=\frac{1}{2}\left(-\ii\frac{\partial}{\partial x}-\frac{B}{2}y\right)^2+\frac{1}{2}\left(-\ii\frac{\partial}{\partial y}+\frac{B}{2}x\right)^2+V(x,y)
\end{equation}
where $V(\cdot,\cdot)=V(\cdot+1,\cdot)=V(\cdot,\cdot+1)$ is a $\Z^2$-periodic potential.   
However, a direct analysis of its fine properties is extremely difficult and one needs resorting to  simpler \emph{effective models} hoping to capture (some of) the
main physical features in suitable physical regimes, such as
for example, in the limit of a weak or strong magnetic field.

In the limit of a strong magnetic field, $B\gg1$, it is well known (cf. \cite{wilkinson-87,bellissard-89,helfer-sjostrand-89,denittis-10,denittis-panati-10})
that the physics of the IQHE, described in full by the Hamiltonian \eqref{eq_000}, is quite well approximated by the effective  operator $H_1^\theta=D_\theta+C$ acting as
\begin{equation}\label{eq_001}
(H_1^\theta\psi)(x)= \psi(x-\theta)+\psi(x+\theta) + 2\cos(2\pi x)\psi(x) ,
\end{equation}
on the Hilbert space $L^2(\R,\dd x)$. Here $\theta:= \nicefrac{1}{B}$ and 
\begin{equation}\label{eq_001bis}
(D_\theta\psi)(x):=\psi(x-\theta)+\psi(x+\theta) , \qquad  (C\psi)(x):= 2\cos(2\pi x)\psi(x) .
\end{equation}

The computation of the spectrum of $H_1^\theta$ is an old problem dating back to the pioneering work of D.R. Hofstadter  \cite{hofstadter-76}.
The collection of the spectra $\sigma(H_1^\theta)$, when $\theta$ varies from $0$ to $1$, results in a two-dimensional fractal diagram known as \emph{quantum butterfly}. When the parameter $\theta$ is rational, that is $\theta=\nicefrac{M}{N}$ (with $M,N$ coprime), the spectrum of $H_1^\theta$ in \eqref{eq_001}
is made of $N$ energy bands if $N$ is odd or  $N-1$ if $N$ is even, respectively. One has $N+1$ gaps (for $N$ odd) or $N$ gaps (for $N$ even) if one includes in the computation the \emph{inf-gap} (i.e. the unbounded gap from $-\infty$ to the minimum of the spectrum) and the \emph{sup-gap} (i.e. the unbounded gap from  the maximum of the spectrum to $+\infty$). 

To each gap $g$ one associates a spectral projection $P_g$
with the convention that $P_0=0$ for the inf-gap $g=0$ and $P_\text{max}=\num{I}$ for the sup-gap $g=N_\text{max}$ with $N_\text{max}=N-1$ or $N_\text{max}=N$ according to whether $N$ is odd or even. 
As usual, the projection $P_g$ of the Hamiltonian $H_1^\theta$ is defined by the spectral subset  $I_g:=[\varepsilon_0,\varepsilon_g]\cap\sigma(H_1^\theta)$ with $\varepsilon_0$ any real number $-\infty<\varepsilon_0<\text{min}\ \sigma(H_1^\theta)$ and $\varepsilon_g$ any real number in the gap $g$. Since $\varepsilon_0,\varepsilon_g\in\R\setminus \sigma(H_1^\theta)$, there exists a 
closed rectifiable path  $\Lambda\subset\C$
intersecting the real axis in  $\varepsilon_0$ and $\varepsilon_g$ and enclosing the interval $I_g$.
The projection $P_g$ is  defined via holomorphic functional calculus using the \emph{Riesz formula}
\begin{equation}\label{rf}
P_g:=\frac{1}{\ii 2\pi}\oint_\Lambda(\lambda\num{I}-H_1^\theta)^{-1}\ \dd \lambda .
\end{equation}

The Hall conductance associated to  the energy spectrum up to the gap $g$ 
is related to the projection $P_g$ via the \emph{Kubo formula} (linear response theory; cf. \cite{morandi-88,Bellissard-Baldes-Elst-94}). Its value is an integer number $t_g$ (in units of ${e^2}/{h}$)  fulfilling the Diophantine equations
\begin{equation}\label{eq_002}
N\ t_g+M\ s_g=d_g\qquad\qquad g=0,\ldots,N_\text{max}.
\end{equation}
The integer $d_g$ in the right-hand side of  \eqref{eq_002} coincides with the labeling of the gap when $N$ is odd, i.e.  $d_g=g$ for $N$ odd. When $N$ is even one has $d_g=g$ if $0\leqslant g\leqslant {N}/{2}-1$ and $d_g=g+1$ if ${N}/{2}\leqslant g\leqslant N_\text{max}=N-1$.
The second integer $s_g$ in  \eqref{eq_002} is commonly interpreted  (cf. \cite{Thouless-Kohmoto-Nightingale-Nijs-82,avron-04} among others) as the Hall conductance associated to the energy spectrum up to the gap $g$ but in the opposite limit of a  weak magnetic field ($B\ll1$).
The numbers  $s_0,\ldots,s_{N_\text{max}}$ are  subjected to the constraint inequalities 
\begin{equation}\label{eq_002bis}
2\, |s_g|<N.
\end{equation}

For any possible $0\leqslant g\leqslant N_\text{max}$, the corresponding equation in \eqref{eq_002} is solved by  infinite pairs $(t_g,s_g)\in\Z^2$. Upon imposing the constraint \eqref{eq_002bis} the solution 
$(t_g,s_g)$ is unique,  provided the following convention is made:

\noindent
{\bf Rationality convention}
\emph{
When $\theta\in\Q$ its representative is fixed as $\theta=\nicefrac{M}{N}$ with $M\in\Z$, $N\in\N\setminus\{0\}$ and $M,N$  coprime, i.e. the greatest common divisor $\text{\upshape g.c.d}(N,M)=1$. 
}

\goodbreak
\medskip

As mentioned, equations \eqref{eq_002} (with the constraint \eqref{eq_002bis}) were established for the first time in \cite{Thouless-Kohmoto-Nightingale-Nijs-82}  and are referred to as \emph{TKNN-equations}. The early insight of considering the integers $s_g$ and $t_g$ as Chern numbers of suitable vector bundles, while not really taken seriously in the many works devoted to a rigorous derivation of \eqref{eq_002} that have appeared in the last thirty years (cf. \cite{sterda-82,macdonald-84,dana-avron-zak-85,avron-yaffe-86} among many others), 
has been given a recent rigorous proof \cite{denittis-10}. The emerging geometric structure comes from a procedure of \virg{bundle representation}
induced by the existence of a family of symmetries for the system. Once the geometric structure is established, the link between spectral quantities and topological invariants follows from  standard ideas as in \cite{simon-83,avron-seiler-simon-83}.
An important consequence of the intrinsic geometric approach, is that the range of validity of the TKNN-equations \eqref{eq_002} is extended
to a large class of operators (indeed a $C^\ast$-algebra) containing $H_1^\theta$. 

This $C^\ast$-algebra is none other that the rational rotation algebra $\sss{A}_{\nicefrac{M}{N}}$ generated by unitary operators  $u,v$ commuting up to a phase, $uv=\expo{\ii 2\pi\nicefrac{M}{N}}\ vu$.
As we shall see in details below, the operator $H_1^\theta$ in \eqref{eq_001} is the image, via a faithful
representation on the Hilbert space $L^2(\R,\dd x)$, of a \virg{universal operator}
$ h_\theta=u+u^\ast+v+v^\ast $.
Different representations of the algebra (named \emph{$(q,r)$-Weyl representations} below) will then lead to operators $H^\theta_{q,r}$ (on suitable Hilbert spaces) which, as shown in Sect.~\ref{sec_int_qWeyl}, all
have all the same spectrum or, in other words, they are \emph{isospectral}. 
Furthermore, 
their spectral projections will be the image, via the representation, of projectors $p$ in the algebra $\sss{A}_{\nicefrac{M}{N}}$. In turn, each of this projection determines a vector bundle 
${L}_{q,r}({p})$ over the torus $\num{T}^2$. For these bundles we have the following result.
\begin{teo}\label{teo:new1}
For any projection $p\in\text{\upshape Proj}(\sss{A}_{\nicefrac{M}{N}})$,
the vector bundle ${L}_{q,r}({p})\to\num{T}^2$ has 
(first) Chern number $C_{q,r}({p}):=C_1({L}_{q,r}({p}))$ given by the formula
\begin{equation}\label{eq:new1}
C_{q,r}({p})= q\left[\nint({p})+\left(\frac{M}{N}-\frac{r}{q}\right)\ncC({p})\right].
\end{equation}
\end{teo}
\noindent
Here the noncommutative integral $\ncint({\cdot})$ and the Connes--Chern character  $\ \ncC({\cdot})\ $ are canonically defined for the algebra $\sss{A}_{\nicefrac{M}{N}}$. 
Out of this we get a
\emph{generalized TKNN-equations} for Hall conductances related to the gap structure of the spectrum of the 
operator $H^\theta_{q,r}$,
\begin{equation}\label{eq:new2}
N t_g+(qM-rN) s_g = q \, d_g, \qquad\quad g=0,\ldots,N_\text{max}.
\end{equation}
Now $t_g=C_{q,r}(p_g)$  with $s_g= -\ncC(p_g)$ and $d_g=N\ncint(p_g)$ as before,
the $p_g$ being 
(projections in $\sss{A}_{\nicefrac{M}{N}}$ corresponding to) the spectral projections of the 
operator $H^\theta_{q,r}$.
These equations reduce to the starting TKNN-equations \eqref{eq_002} for $q=1$ and $r=0$.

\goodbreak
\medskip

As a particular example of the above, we get the Dirac-like Hamiltonian
\begin{equation}\label{eq_003}
H^\theta_{2,1}:=\left(
\begin{array}{cc}
C &  D_{\theta-\frac{1}{2}}\\ 
D_{\theta-\frac{1}{2}} & -C
\end{array}\right)
\end{equation}
acting on the Hilbert space $L^2(\R,\dd x)\otimes\C^2$. Here $D_{\theta-\frac{1}{2}}$ and $C$ are again given by \eqref{eq_001bis}. 
This operators comes from a mathematical description of physics model for the IQHE on graphene 
(cf. \cite{bellissard-kreft-seiler-91, hatsugai-fukui-aoki-06, sato-tobe-kohmoto-08}). 
Operators like $H_{2,1}^\theta$  
can also be used  to describe effective models for electrons  interacting with the periodic structure of a crystal
through a periodic (internal) magnetic field and subjected to the action of an external strong magnetic field  \cite[Thm.~4.4.12]{denittis-10}. The generalized TKNN-equation above is then 
\begin{equation} 
N t_g + (2M-N)s_g = 2 d_g, \qquad\quad g=0,\ldots,N_\text{max}. 
\end{equation}
This  diophantine equation bears similarities with an analogous one found in where \cite{sato-tobe-kohmoto-08}.

We stress that models like \eqref{eq_000} or \eqref{eq_003} explains only the quantization of the transverse conductance (geometric effect). This is only one of the surprising aspects of the QHE.  
An interesting statistical aspect showed by the QHE is the presence of the \emph{plateaux} which is related to the presence of disorder \cite{Bellissard-Baldes-Elst-94,germinet-klein-schenker-09}. These aspects are out of the scope of the present  work, devoted to an analysis of the geometry emerging from models for the QHE.

The paper is organized as follow. 
Sect~\ref{se:mainres} is devoted to the presentation of the main results of this paper. 
In 
Sect.~\ref{se:ncgf} we introduce the geometry of the noncommutative torus
with its \emph{$(q,r)$-Weyl representations} in  Sect.~\ref{sec_int_qWeyl}, while in Sect.~\ref{sec_cc_NC} we recall the noncommutative integral and the natural derivations on it.
In Sect.~\ref{sec_int_NC} we translate the TKNN-equations \eqref{eq_002} in the geometrical language of the noncommutative torus. 
Sect.~\ref{sec_int_main} is devoted to expose our main results: Theorem \ref{teo_main_1} which proves the existence of a family of \emph{bundle representations} (parametrized by $q$ and $r$)  for the (rational) noncommutative torus and the (main lines of the) proof of Theorem \ref{teo:new1} which states the generalized version of the TKNN-equations for the $(q,r)$-Weyl representations. Some consequences, as the possibility to extend the results for the irrational case are explored in Sect.~\ref{sec_int_com_rk}.
Sect.~\ref{sec_BF} contains the proof of  Theorem \ref{teo_main_1}. In Sect.~\ref{sub_sec_3.1}, on the base of a generalized version of the Bloch-Floquet theory, we provide a \emph{direct integral representation} for the    $(q,r)$-Weyl representations (in the rational case). 
In Sect.~\ref{sec_loc_triv} we show that a geometric structure (a vector bundle with connection) 
emerges in a straightforward way from the direct integral decomposition. In  Sect.~\ref{sec_bund_rep1} we finally prove that any   $(q,r)$-Weyl representation produces a   bundle representation for the noncommutative torus algebra.
Sect.~\ref{sec_gem_dual} is devoted to the technicalities needed for the proof of  Theorem \ref{teo:new1}. In Sect.~\ref{sec_geoduality} we introduce a \emph{reference} bundle representation
whose geometry comes naturally and canonically from the geometry of the noncommutative torus. 
This {reference} bundle representation is coupled with the $(q,r)$-Weyl representations by means of a \emph{geometric duality formula} (Theorems  \ref{teo_main_3}). The proof of such a formula is the content of  Sect.~\ref{sec_geoduality}. In Sect.~\ref{sec_RBR} we finally derive the generalized TKNN-equations as a simple consequence of the geometric duality. 
In App.~\ref{se:ccn} we relegate some computation of Chern numbers using charts and transition functions.

\goodbreak
\medskip

\noindent{\bf Acknowledgments:} G.D. would like to thank G. Dell'Antonio and G. Panati for their constant advise and encouragement, and J. Kellendonk for many stimulating discussions. {We are grateful to M. Rieffel for useful remarks}. G.D. is supported by the grant ANR-08-BLAN-0261-01. G.L. was partially supported by the Italian Project \virg{Cofin08 -- Noncommutative Geometry, Quantum Groups and Applications}.

\section{Main results: generalized TKNN-equations}\label{se:mainres}

The relevance of the \emph{(ir)rational rotation algebra} \cite{rieffel-81} or \emph{noncommutative torus} \cite{connes-80}, for the quantum Hall effect and in particular for the study of operators like the one in \eqref{eq_001} is well established starting from the early work \cite{bellissard-89} to the more recent ones \cite{marcolli-mathai-06}. 

\subsection{The noncommutative geometry framework}\label{se:ncgf}

The noncommutative torus (NCT) is perhaps the best known example of a noncommutative manifold. Here we briefly recall some results that we shall need below, using the 
compendiums \cite{boca-01} and \cite{gracia-varilly-figueroa-01} as main sources.

The $C^\ast$-algebra of the NCT (or \emph{NCT-algebra}) is defined in a universal way  starting from two elements $u$ and $v$ which are unitary with respect to 
an involution $\ast$, i.e. $u^\ast=u^{-1}$ and $v^\ast=v^{-1}$
and which commute up to a phase:  
\begin{equation}\label{nct0}
uv=\expo{\ii 2\pi\theta}\ vu , \qquad \mathrm{with} \quad \theta\in\R .
\end{equation}
The space $\sss{L}_\theta$ of finite complex linear combinations  of the monomials $u^nm^m$, with $n,m\in\Z$ has a natural structure of a unital $\ast$-algebra
with unit $u^0=\num{I}=v^0$. The NCT-algebra $\sss{A}_\theta$ with \emph{deformation parameter} $\theta$, 
is the $C^\ast$-algebra obtained as the closure of $\sss{L}_\theta$ with respect to the universal norm, 
$\|a\|:=\sup\left\{\|\pi( a )\|_{\bbb{B}(\sss{H})}\right\}$, 
where the supremum  is taken over all the  $\ast$-representations  $\pi:\sss{L}_\theta\to \bbb{B}(\sss{H})$, with 
$\bbb{B}(\sss{H})$ denoting the $C^\ast$-algebra of bounded operators on the Hilbert space
$\sss{H}$. 
When $\theta\in\Q$ the NCT-algebra $\sss{A}_\theta$ is called \emph{rational}. 

The universal behavior of the NCT-algebra is stated as follows. Let ${U}$ and ${V}$ be a
pair of unitary operators  acting on a Hilbert space and $\sss{H}$ such that $UV=\expo{\ii 2\pi \theta}VU$
and denote with $C^\ast(U,V)\subset\bbb{B}(\sss{H})$  the $C^
\ast$-algebra generated by $U$ and $V$. Universality means that  the mapping $\pi(u)=U$, $\pi(v)=V$ extends
algebraically to a representation $\pi:\sss{A}_\theta\to C^\ast(U,V)$ of $\sss{A}_\theta$ which is \emph{surjective}.
The universality property of  $\sss{A}_\theta$ also entails the fact that its algebraic structure 
does not depend on the particular choice of a pair of generators $(u,v)$ (a \emph{frame}). Any other choice of a frame, i.e. any other pair of unitaries $(u',v')$ in $\sss{A}_\theta$ such that $u'v'=\expo{\ii 2\pi\theta}\ v'u'$ 
provides an equivalent system of generators for $\sss{A}_\theta$. 
Thus, there is no canonical choice for the system of generators of $\sss{A}_\theta$. In the rest of the paper we  refer to the algebra $\sss{A}_\theta$ assuming an \virg{a priori} fixed choice  $(u,v)$ of a frame.

\goodbreak
\medskip

The link between the NCT-algebra and the Hamiltonian \eqref{eq_001} becomes apparent upon considering the unitary operators
$T_1$ and ${T^\theta_2}$ on $L^2(\R,\dd x)$ defined by
\begin{equation}\label{eq_004}
(T_1 \psi)(x) :=\expo{\ii 2\pi x}\psi\left(x\right)   \qquad \mathrm{and} \qquad
(T^\theta_2 \psi)(x) :=\psi\left(x-\theta\right)
\end{equation}
which are readily seen to obey the relation $T_1T^\theta_2=\expo{\ii 2\pi \theta}\ T^\theta_2T_1$. 
From the universality, the mapping  $\Pi_1(u):=T_1$ and 
$\Pi_1(v):={T^\theta_2}$ extends to a surjective representation  $\Pi_1:\sss{A}_\theta\to C^\ast(T_1,{T^\theta_2})$, named the \emph{1-dimensional Weyl representation}. It turns out that, independently of $\theta\in\R$, the  representation $\Pi_1$ is faithful. This claim is trivial when $\theta\in\R\setminus\Q$, indeed any representation of the irrational NCT-algebra is automatically faithful 
\cite[Thm.~1.10]{boca-01}.
On the other hand, when $\theta=\nicefrac{M}{N}$,
a simple criterion to check 
the faithfulness  is to  show that the commutative $C^\ast$-algebra $C^\ast((T_1)^N,(T_2^{\theta})^N)$ is $\ast$-isomorphic to the torus algebra $C(\num{T}^2)$ \cite[Prop.~1.11]{boca-01},  which is equivalent (in view of the Gel'fand isomorphism) to prove that the \emph{joint spectrum} of $(T_2)^N$ and $(T_2^{\theta})^N$ is $\num{T}^2$. This last claim follows from a direct computation of the simultaneous (generalized) eigenvectors of $(T_2)^N$ and $(T_2^{\theta})^N$ (cf. \cite[Sect.~5.1.3]{denittis-10}). 

Upon observing that $T_1+{T_1}^{-1}=C$ and ${T^\theta_2}+{T^{-\theta}_2}=D_\theta$ of \eqref{eq_001bis}, 
it follows from \eqref{eq_001} that $H_1^\theta=\Pi_1(h_\theta)$ with the universal \emph{Hofstadter operator} $h_\theta\in\sss{A}_\theta$ defined by
\begin{equation}\label{eq_005}
h_\theta:=u+u^\ast+v+v^\ast.
\end{equation}
The faithfulness of the representation $\Pi_1$ implies that the spectrum of $ h_\theta$ as element of the $C^\ast$-algebra 
$\sss{A}_\theta$ equates the spectrum of $H_1^\theta$ as bounded operator on $L^2(\R,\dd x)$. 

Let $C^\ast(H_1^\theta)\subset\bbb{B}(\sss{H}_1)$ be the $C^\ast$-algebra generated by the operator $H_1^\theta$. Clearly $C^\ast(H_1^\theta)\subset C^\ast(T_1,T_2^\theta)$. Being $C^\ast(H_1^\theta)$ closed with respect to the holomorphic functional calculus, it follows that $P_g\in C^\ast(H_1^\theta)$ where
$P_g$ denotes the
spectral projection  of $H_1^\theta$ defined by the Riesz formula \eqref{rf}. The faithfulness of the representation $\Pi_1$ implies that there exists a unique \emph{projection} $p_g\in\sss{A}_\theta$ such that $\Pi_1(p_g)=P_g$. Thus, any spectral projection of $H_1^\theta$ related to a gap is the image via $\Pi_1$ of a (unique) universal projection in the NCT-algebra $\sss{A}_\theta$.
As mentioned in Sect.~\ref{se:intro} 
the integers  $t_g$, $s_g$ and $d_g$ which appear in the TKNN-equations \eqref{eq_002} are related to the spectral projection $P_g$. The above considerations lead to conclude  that  these integers are related to the universal projection $p_g$ in a way that we shall illustrate in Sect.~\ref{sec_int_NC}.

\subsection{$q$-dimensional  $r$-twisted Weyl representations}\label{sec_int_qWeyl}

The representation induced by \eqref{eq_004} can be generalized to higher dimensional versions. These were first introduced in \cite{connes-rieffel-87} and used for Yang--Mills connections on the NCT.
For any $q\in\N\setminus\{0\}$  one sets 
$$
\sss{H}_q:=L^2(\R,\dd x)\otimes\C^q\simeq L^2(\R,\dd x;\C^q)
$$ 
and define  two (one parameter families of) $q\times q$ complex matrices 
\begin{equation}\label{eq_005bis}
\num{U}_q(\lambda):=\lambda\left(
\begin{array}{lllc}
1 & 0 &  \ldots &  0 \\ 
0 & \expo{\ii  \frac{2\pi}{q}} &   \ldots &  0 \\ 
\vdots & \vdots &  \ddots &  \vdots\\ 
0 & 0 &   \ldots & \expo{\ii  \frac{2\pi(q-1)}{q}} 
                       \end{array}\right),
                       \  \ \ \ \num{V}_q(\lambda):=\left(
\begin{array}{llllll}
0 &     \ldots & 0 & \lambda \\ 
1 &     \ldots & 0 & 0 \\ 
\vdots &    \ddots & \vdots & \vdots \\ 
0 &     \ldots & 1 & 0
                       \end{array}\right)
\end{equation}
with $\lambda\in{\mathbb S}:=\{z\in\C \ :\ |z|=1\}$. 
For any two $\lambda,\lambda'\in{\mathbb S}$, a simple computation yields 
\begin{equation}\label{ecr}
\num{U}_q(\lambda)\ \num{V}_q(\lambda')=\expo{\ii 2\pi \frac{1}{q}}\ \num{V}_q(\lambda')\ \num{U}_q(\lambda) .
\end{equation}
It is also easy to compute that 
$(\num{U}_q(\lambda))^q=\lambda^q \num{I}$ and $(\num{V}_q(\lambda))^q=\lambda \num{I}$, for any $\lambda \in {\mathbb S}$.
We shall use the shorthand notation  
\begin{equation}\label{shn}
{\num{U}}_q:=\num{U}_q(1), \qquad  \textup{and} \qquad {\num{V}}_q:=\num{V}_q(1) . 
\end{equation}
Then, 
for any $r\in\{\pm1,\ldots,\pm(q-1)\}$ coprime with respect to $q$ (i.e. $\text{\upshape g.c.d}(q,r)=1)$ one defines a pair  of unitary operators on  $\sss{H}_q$ by
\begin{equation}\label{eq_006}
U_q:=T_1\otimes\num{U}_q, \ \ \ \ \ \  \ \ V^\theta_{q,r}:=T_2^{\epsilon}\otimes\num{V}_q^r, \ \ \ \ \ \  \ \ \epsilon(\theta,q,r):=\theta-\frac{r}{q} ,
\end{equation}
with $T_1$ and $T^\epsilon_2$ given by \eqref{eq_004}.
Let $\{e_0,\ldots,e_{q-1}\}$ be the canonical basis of $\C^q$.
The set of vectors $\Psi(\cdot):=\sum_{\ell=0}^{q-1}\psi(\cdot\ ;\ell)\otimes e_\ell$ with $\psi(\cdot\ ;\ell)\in L^2(\num{R},\dd x)$ for any $\ell=0,\ldots,q-1$, 
is dense in $\sss{H}_q$. The action of $U_q$ and $V^\theta_{q,r}$ on the components of $\Psi(\cdot)$ is found to be given by
\begin{equation}\label{eq_007}
(U_q\psi)(x;\ell)=\expo{\ii 2\pi\left(x+\frac{\ell}{q}\right)}\psi\left(x;\ell\right) \qquad \mathrm{and} \qquad 
(V^\theta_{q,r}\psi)(x;\ell)=\psi\left(x-\theta+\frac{r}{q};[\ell-r]_q\right)
\end{equation}
where $[\cdot]_q$ denotes the class modulo $q$. Again, one readily shows that 
$U_qV^\theta_{q,r}=\expo{\ii 2\pi \theta}\ V^\theta_{q,r}U_q$.

The mapping $\Pi_{q,r}(u):=U_q$ and $\Pi_{q,r}(v):=V^\theta_{q,r}$ 
extends to a surjective representation  $\Pi_{q,r}:\sss{A}_\theta\to C^\ast(U_q,V^\theta_{q,r})$, named the \emph{$q$-dimensional $r$-twisted Weyl representation}
or more succinctly the \emph{$(q,r)$-Weyl representation}.
As it is the case for the representation $\Pi_1:=\Pi_{1,0}$ defined by \eqref{eq_004},  
the representations $\Pi_{q,r}$ is faithful for any $\theta\in\R$.

\begin{rk}[$0$-twisted representations]\label{rk_01}
The case of a $q$-dimensional Weyl representation with twisting $r=0$ is quite trivial. Indeed $\Pi_{q,0}$ reduces, up to a unitary equivalence, to  $q$ copies of the $1$-dimensional representation $\Pi_1$. Equation \eqref{eq_006} shows that $V_{q,0}$ coincides with the operator
$\oplus_{\ell=0}^{q}T^\theta_2$ on the Hilbert space $\oplus_{\ell=0}^{q}\sss{H}_1$. Also $U_q$ defines an operator on $\oplus_{\ell=0}^{q}\sss{H}_1$ which acts as $\oplus_{\ell=0}^{q}\expo{\ii 2\pi\frac{\ell}{q}} T_1$ with $\omega=\expo{\ii \frac{2\pi}{q}}$.
However, $U_q$  is not a sum of $q$ copies of a single operator. Consider the unitary operator $R:=\oplus_{\ell=0}^{q}{T_2}^{\nicefrac{\ell}{q}}$. A simple computation shows that $R\ U_q\ R^{-1}=\oplus_{\ell=0}^{q}{T_1}$
and $[R,V_{q,0}]=0$, namely  $R\ \Pi_{q,0}(\sss{A}_\theta)\ R^{-1}=\oplus_{\ell=0}^{q}\Pi_{1}(\sss{A}_\theta)$.\\ 
$\phantom{a}$\hfill $\blacklozenge\lozenge$
\end{rk}

A comparison between the Dirac-like operator $H^\theta_{2,1}$ defined by \eqref{eq_003}, the  Hofstadter operator $h_\theta$ defined by \eqref{eq_005} and the unitaries $U_2$ and $V^\theta_{2,1}$ given by \eqref{eq_006} shows that $H^\theta_{2,1}=\Pi_{1,2}(h_\theta)$. 
The faithfulness of the representation $\Pi_{2,1}$ implies isospectrality between $H^\theta_{2,1}$ and 
$h_\theta$. From the faithfulness of the representation $\Pi_1$, one in turn infers that $\sigma(H^\theta_{1})=\sigma(H^\theta_{2,1})$. 
In particular, this means that $H^\theta_{1}$ and $H^\theta_{2,1}$ have the same system of gaps in the spectrum and so the same family of spectral projections in the gap. As explained at the end of Sect.~\ref{se:ncgf}, the spectral projections into the gaps can be realized as the representation of a (unique) universal projection in $\sss{A}_\theta$. These considerations extend to each   Weyl representation and to any (self-adjoint) element in $\sss{A}_\theta$. In other words, let $h_\theta\in\sss{A}_\theta$ be any self-adjoint universal operator, not necessarily the Hofstadter operator, then for any $(q,r)$-Weyl representation: 
\begin{itemize}
\item $\sigma(H^\theta_{q,r})=\sigma_{\sss{A}_\theta}(h_\theta)$ where $H^\theta_{q,r}:=\Pi_{q,r}(h_\theta)$ is the image in the representation, and $\sigma_{\sss{A}_\theta}(\cdot)$ denotes the (algebraic) spectrum in 
the $C^\ast$-algebra $\sss{A}_\theta$;
\item let $p\in C^\ast(h_\theta)\subset\sss{A}_\theta$ be a universal projection, then $\Pi_{q,r}(p)$ is a spectral projection of  $H^\theta_{q,r}$ associated with some subset of the spectrum disconnected by gaps from the rest of the  spectrum.
\end{itemize}
 
The representations $\Pi_{q,r}$ 
are the main object of interest of our paper. 
We shall use them in Sect.~\ref{sec_int_main} to derive a system of TKNN-equations for any $(q,r)$-Weyl representation.

\subsection{The integral and the character}\label{sec_cc_NC}

In order to proceed we need additional structures on the NCT-algebra: 
these are a natural integral and natural derivations. Let $(u,v)$ be a (fixed) system of generators for $\sss{A}_\theta$. The linear map 
$\ncint\ :\sss{A}_\theta\to\C$ defined on monomials by 
$$\nint( u^n  v^m):=\delta_{n,0}\ \delta_{m,0}$$ 
extends to all $\sss{A}_\theta$ by linearity. It is indeed a faithful state on $\sss{A}_\theta$ with the trace property $\ncint( a b - b a )=0$ for any $ a, b \in\sss{A}_\theta$ (cf.
\cite{boca-01}). We refer to $\ncint\ $ as the \emph{noncommutative integral} over $\sss{A}_\theta$.
In general the definition of $\ncint\ $ is not canonical since it is subordinate to the choice of a system of generators $(u,v)$. It is canonical for irrational $\theta\in\R\setminus\Q$ since in this case there exists a unique tracial state on $\sss{A}_\theta$.

The  \emph{derivations} $\ncpartial_j:\sss{A}_\theta\to\sss{A}_\theta$, $j=1,2$, are defined on the monomials by the equations
\begin{equation}\label{eq_008}
\ncpartial_1(u^nv^m)=\ii 2\pi\ n\ u^nv^m,\ \ \ \ \ \ \ \ \ \ \  \ncpartial_2(u^nv^m)=\ii 2\pi\ m\ u^nv^m ,
\end{equation}
and extended by linearity and Leibniz rule. Equation \eqref{eq_008} shows that  $\ncpartial_1$ and $\ncpartial_2$ commute and are unbounded on $\sss{A}_\theta$. Their common maximal invariant domain 
is called the \emph{smooth NCT-algebra} and is denoted with $\sss{A}_\theta^\infty$.
The {smooth algebra} $\sss{A}_\theta^\infty$ is a dense unital $\ast$-algebra in $\sss{A}_\theta$ stable under the holomorphic functional calculus, i.e. it is a Fr\'echet unital pre-$C^\ast$-algebra (cf. \cite[Def.~12.6]{gracia-varilly-figueroa-01}).
Elements in $\sss{A}_\theta^\infty$ are power series of the form
\begin{equation}\label{2ta}
a =\sum_{(n,m) \in \Z^2} a_{n,m}~u^n\,v^m~,
\end{equation}
with $\{a_{n,m}\}\in S(\Z^2)$ a complex-valued Schwartz function
on $\Z^2$. This means that the sequence of numbers $\{a_{n,m} \in
\C~:~ (n,m) \in\Z^2 \}$ decreases rapidly at \virg{infinity}, i.e. 
one has bounded semi-norms
$
\|a\|_k = \sup_{(n,m)\in \Z^2} ~|a_{n,m}|\,\big(1+ |n|+|m| \big)^k < \infty 
$, for any $k\in\N\setminus\{0\}$

To lighten notation, in the following we shall use the symbol $\sss{A}_\theta$ also for the smooth subalgebra $\sss{A}^\infty_\theta$ always being evident when we are dealing with smooth elements.

We shall denote by $\text{\upshape Proj}(\sss{A}_\theta)$ the space of all projections in $\sss{A}_\theta$. For any (smooth) element of $\text{\upshape Proj}(\sss{A}_\theta)$ one defines the \emph{Connes--Chern character} which, for the NCT-algebra is made up of two components. 
Firstly (or better secondly) there is a \virg{$2$-form} resulting into a map $\ncC\ :\text{{\upshape Proj}}(\sss{A}_\theta)\to\Z$ defined by the Connes formula \cite{connes-80}:\begin{equation}\label{eq_009}
\ncC(p):=\frac{1}{\ii 2\pi}\nint p \big(\ncpartial_1(p)
\ncpartial_2(p)-\ncpartial_2(p)\ncpartial_1(p) \big) .
\end{equation}

The other piece is a \virg{$0$-form}  or a rank function. 
For rational values of the deformation parameter $\theta=\nicefrac{M}{N}$ 
we take the \emph{rank-function} 
$\text{Rk}$ to be  normalized  as 
\begin{equation}\label{tra}
\text{Rk}(\cdot) :=N\nint (\cdot) . 
\end{equation}
Arguments for this being a good definition are that $\text{\upshape Rk}:\text{\upshape Proj}(\sss{A}_{\nicefrac{M}{N}})\to\{0,1,\ldots,N\}$ \cite[Cor.~1.22]{boca-01}  
and that faithfulness of the noncommutative integral implies 
$\text{\upshape Rk}({p})=0$ if and only if ${p}=0$  while $\text{\upshape Rk}({p})=N$ if and only if  
$p=\num{I}$.

\subsection{A noncommutative geometric look at the TKNN-equations}\label{sec_int_NC}

Let us come back to the analysis of the TKNN-equations in \eqref{eq_002}.
The first relevant observation comes from the fact that the $C^\ast$-algebra $\Pi_1(\sss{A}_{\nicefrac{M}{N}})\subset\bbb{B}(L^2(\R,\dd x))$ admits a bundle decomposition over a rank $N$ Hermitian vector bundle $E_{N,1}\to\num{T}^2$ , as we shall see in generality in Sect.~\ref{sec_int_main}. In view of the Serre-Swann theorem (cf. \cite[Thm.~2.10]{gracia-varilly-figueroa-01}) which describes vector bundles via their modules of sections, the spectral projection $P_g$ in \eqref{rf} of the Hamiltonian $H_1^\theta$ defines a vector subbundle $L(P_g)\subset E_{N,1}$. The first Chern class of $L(P_g)$ is the only non trivial Chern class, due to the low dimensionality of the base manifold. The related (first) Chern number $C_1(L(P_g))$ measures the degree of non triviality of the  vector bundle $L(P_g)$.
The geometric interpretation of the integer $t_g$ in \eqref{eq_002} is none other than the equality $t_g=C_1(L(P_g))$. To make explicit the dependence of $L(P_g)$ and $C_1(L(P_g))$ on the abstract projection $p_g$ via the representation $\Pi_1$ we use the more concise  notation $L_1(p_g):=L(P_g)$ and $C_1(p_g):=C_1(L(P_g))$. Thus we have that
$$
t_g=C_1(p_g) .
$$
The next step consists in showing that $s_g= -\ncC(p_g)$ and $d_g=\text{Rk}(p_g)$ with these maps defined as above in \eqref{eq_009} and \eqref{tra}. In particular, the first equality means that the Hall conductance in the weak magnetic field regime is given (up to a sign) by the Connes--Chern character. This has been proved in   \cite{denittis-10} (cf. also \cite{denittis-panati-10}), culminating in the Prop.~5.2.2 there, 
to which we refer for all details.  Here we only mentions the relevant points of the two-step proof. In the first step, leading to $s_g= -\ncC(p_g)$, one shows that  in the adiabatic limit $B\ll 1$, the
Schr\"{o}dinger operator  for the magnetic Bloch electron
\eqref{eq_000} in a suitable range of energy is (asymptotically)
unitarily equivalent to the effective operator
\begin{equation}\label{eq_new_1}
H_0^{\theta}:=K_{\theta}+{K_{\theta}}^{-1}+G_{\theta}+{G_{\theta}}^{-1},
\end{equation}
where now $\theta:=B$ (recall that in the strong field limit
$\theta=B^{-1}$). The unitary operators $K_{\theta}$ and
$G_{\theta}$ act on the Hilbert space $\sss{H}_0:=\ell^2(\Z^2)$
according to
\begin{equation}\label{eq_new_2}
(K_{\theta}\xi)_{n_1,n_2}:=\expo{\ii \pi n_2\theta}\
\xi_{n_1+1,n_2},\qquad (G_{\theta}\xi)_{n_1,n_2}:=\expo{-\ii
\pi n_1\theta}\ \xi_{n_1,n_2+1}
\end{equation}
where $\xi_{n_1,n_2}$, with $n=(n_1,n_2)\in\Z^2$ form an orthonormal basis for the Hilbert space 
$\sss{H}_0$. A
straightforward  computation shows that 
$K_{\theta} G_{\theta}=\expo{\ii 2\pi \theta} G_{\theta} K_{\theta}$, the NCT-algebra relation. 
Then the
mapping $\Pi_0(u):=K_{\theta}$ and $\Pi_0(v):=G_{\theta}$ extends to a
surjective representation
$\Pi_0:\sss{A}_{\theta}\to C^\ast(K_{\theta},G_{\theta})\subset {\sss{H}_0}$ which turns out
to be (unitarily equivalent to) the GNS representation of the
NCT-algebra $\sss{A}_{\theta}$ associated to the noncommutative
integral $\ncint\ $. This means that
$H_0^{\theta}=\Pi_0(h_{\theta})$, namely the effective model
\eqref{eq_new_1} is (unitarily equivalent to) the GNS realization of
the universal Hofstadter operator $h_{\theta}\in\sss{A}_{\theta}$
defined by \eqref{eq_005}. Now, any spectral projection $p_g$ of
$h_{\theta}$ defines a spectral projection $\Pi_0(p_g)$ of
$H^0_{\theta}$ and 
the weak magnetic field Hall conductance associated to $\Pi_0(p_g)$
and  denoted by
$s_g$, turns out to be (via the Kubo formula) the first Chern number
of a suitable line bundle $L(\Pi_0(p_g))$ associated
to the universal projection $p_g$ via the representation $\Pi_0$. The
second step of the proof, the possibility to relate $\Pi_0(p_g)$ to a vector bundle,
depends on the fact that the $C^\ast$-algebra
$C^\ast(K_{\theta},G_{\theta})$ admits a bundle representation (over a
trivial vector bundle) of the type of
the bundle representation ${\Pi_\text{ref}}$ defined in more generality in
Sect.~\ref{sec_RBR} below and leading to the equality $s_g= -\ncC(p_g)$ being proved
in  general in Proposition~\ref{prop_abst_chern}.  

Collecting all of the above results, we will rewrite the TKNN-equations \eqref{eq_002} as follows:
\begin{equation}\label{eq_010}
C_1(p_g)=\frac{1}{N}\text{Rk}(p_g) +\frac{M}{N}\ncC(p_g)=\nint(p_g) +\frac{M}{N}\ncC(p_g)\\
\qquad\qquad g=0,\ldots,N_\text{max}.\\
\end{equation}
where $p_g$ are the spectral projections of the open gaps of the Hofstadter operator \eqref{eq_005}. With this notation the constraint equation \eqref{eq_002bis} reads
\begin{equation}\label{eq_010bis}
2\, |\!\ncC(p_g)|<N
\end{equation}

One of the consequences of Theorem \ref{teo:new1} is that equation \eqref{eq_010} (which is a special case of equation \eqref{eq:new1}) holds true for any smooth projection:   
\begin{equation}\label{eq_011}
C_1(p)=\nint(p) +\frac{M}{N} \ncC(p) , \qquad \qquad \mathrm{for} \qquad 
p\in\text{Proj}(\sss{A}_{\nicefrac{M}{N}}).
\end{equation}
where $C_1(p):=C_1(L(P))$ is the first Chern number of the vector bundle $L(P)\to\num{T}^2$ associated with  the represented projection $P:=\Pi_1(p)\in\bbb{B}(\sss{H}_1)$ via the Serre-Swan theorem.
This makes explicit the geometric nature and the noncommutative content of the equation \eqref{eq_002}. We will refer to \eqref{eq_011} as the \emph{natural form} for the {TKNN}-equations.
\begin{rk}
The bound  \eqref{eq_010bis} is not valid, in general, for an arbitrary (smooth) projection but it holds only for the spectral projections into the gaps of the Hofstadter operator \cite{Choi-elliott-yui-90}. In principle different operators leads to different bounds and each bound depends on the form of the related operators.
\hfill $\blacklozenge\lozenge$
\end{rk}

\subsection{Generalizing the TKNN-equation}\label{sec_int_main}
We are now ready to present the main results of this paper. The equation \eqref{eq_011} depends on the 1-dimensional Weyl representation $\Pi_1$ used in the definition of the Chern number $C_1(p):=C_1(L(\Pi_1(p)))$.
The first step for its generalization to the higher-dimensional Weyl representations $\Pi_{q,r}$
is to provide a  bundle representation for the $C^\ast$-algebra $\Pi_{q,r}(\sss{A}_{\nicefrac{M}{N}})$. This will then allow to define Chern numbers related to projections in $\sss{A}_{\nicefrac{M}{N}}$.
\begin{teo}[bundle representation]\label{teo_main_1} 
Let $\sss{A}_{\nicefrac{M}{N}}$ be the  rational NCT-algebra,
and let $\Pi_{q,r}:\sss{A}_{\nicefrac{M}{N}}\to \bbb{B}(\sss{H}_q)$ be the $(q,r)$-Weyl representation.
Then the operator algebra $\Pi_{q,r}(\sss{A}_{\nicefrac{M}{N}})$ admits a bundle representation over $\num{T}^2$. This means 
there exists a Hermitian vector bundle ${E}_{N,q}\to\num{T}^2$ and a unitary transform 
$\bbb{F}_{q,r}:\sss{H}_q\to{L^2}({E}_{N,q})$ such that 
$$
\widetilde\Pi_{q,r}(\sss{A}_{\nicefrac{M}{N}}):=\bbb{F}_{q,r}\ \Pi_{q,r}(\sss{A}_{\nicefrac{M}{N}})\ {\bbb{F}_{q,r}}^{-1}\subset\Gamma(\text{\upshape End}({E}_{N,q})) .
$$
The
vector bundle ${E}_{N,q}$  has rank $N$ and (first) Chern number $C_1({E}_{N,q})=q$. 
\end{teo}
The bundle representation of $\sss{A}_{\nicefrac{M}{N}}$ is implemented by the unitary map $\bbb{F}_{q,r}$, called \emph{(generalized) Bloch-Floquet transform} \cite[Chap.~4]{denittis-10}. The explicit recipe for $\bbb{F}_{q,r}$ and the technicalities concerning the proof of Theorem \ref{teo_main_1} are postponed to Sect.~\ref{sec_BF}. 
There, we shall endow the bundle $E_{N,q}$ with a constant curvature connection, the curvature computed to be  
$$ 
K^{(N,q)} =\left(\frac{ 2\pi q}{\ii N}\ \num{I}_N\right) \dd k_1\wedge \dd k_2 .
$$ 
The usual first Chern class, $c_1(E_{N,q}) =\frac{\ii}{2\pi}\text{Tr}_N(K^{(N,q)})=q\ \dd k_1\wedge \dd k_2$, when integrated then yields as corresponding number: 
$$
C_1(E_{N,q})=\int_{\num{T}^2}c_1(E_{N,q})=q .
$$

\begin{rk}\label{rk_fibrep}
The previous theorem asserts that the rational torus algebra $\sss{A}_{\nicefrac{M}{N}}$ admits a discrete family of bundle representations $\widetilde\Pi_{q,r}(\cdot):= \bbb{F}_{q,r}\circ\Pi_{q,r}(\cdot) \circ{\bbb{F}_{q,r}}^{-1}$ each of which specified by a twisting indexed $q$. A characterization of the algebra $\sss{A}_{\nicefrac{M}{N}}$ as the endomorphism algebra of a bundle with (first) Chern number $q$ was already established in \cite[Thm.~3.1]{rieffel-83} albeit 
in a different context and with different techniques. We stress that these are different from the (isomorphic) realization of the rational torus algebra $\sss{A}_{\nicefrac{M}{N}}$ as the algebra of continuous sections of a vector bundle over the two-dimensional torus with typical fiber $\text{Mat}_N(\C)$ and twisting index $M$ (cf. \cite{hoegh-skjelbred-81} or \cite[Prop.~12.2]{gracia-varilly-figueroa-01}).
\hfill $\blacklozenge\lozenge$ 
\end{rk}

The representation of  $\sss{A}_{\nicefrac{M}{N}}$ in Theorem \ref{teo_main_1} comes from a fiber decomposition of the representation Hilbert space $\sss{H}_q$. The unitary equivalence between $\Pi_{q,r}$ and  $\widetilde\Pi_{q,r}$  assures that all spectral information (quantities related to the Hilbertian structure) carried by the representation $\Pi_{q,r}$ are preserved. Vice versa topological quantities, e.g. Chern numbers, emerging out of the bundle representation provide information about the specific Hilbert space representation $\Pi_{q,r}$ and not about the abstract algebraic structure of $\sss{A}_{\nicefrac{M}{N}}$.

\goodbreak
\medskip

Let $P:=\Pi_{q,r}(p)$ be the  $(q,r)$-Weyl representation of a projection ${p}\in\text{Proj}({A}_{\nicefrac{M}{N}})$.  It is unitarily equivalent to the projection $P(\cdot):=\widetilde\Pi_{q,r}(p)\in \Gamma(\text{\upshape End}({E}_{N,q}))$ from Theorem~\ref{teo_main_1}.
In turn, in the spirit of Serre-Swan theorem (cf. \cite[Thm.~2.10]{gracia-varilly-figueroa-01}), the projector $P(\cdot)$ selects a vector subbundle $L_{r,q}(p)\to\num{T}^2$ of the vector bundle $E_{N,q}$. We let $C_{q,r}({p}):=C_1({L}_{q,r}({p}))$ denotes its first Chern number.

Our goal is twofold. On the one hand, we relate the Chern number $C_1({L}_{q,r}({p}))$ to Chern numbers of a \virg{dual} or \virg{reference} bundle leading to a formula generalizing equation \eqref{eq_011}.
Secondly, in a more abstract version, we compute the Chern number $C_{q,r}(p)$ in terms of the noncommutative integral $\ncint$ \,and of the Connes--Chern map $\ncC$\;. Again the resulting formula reducing to the \eqref{eq_011} in the case of the 1-dimensional representation $\Pi_1$. 

\begin{teo}\label{teo_main_3}
For any $p\in\text{{\upshape Proj}}(\sss{A}_{\nicefrac{M}{N}})$ there exists a 
\emph{dual} (or \emph{reference}) vector bundle 
${L}_\text{{\upshape ref}}(p)\to\num{T}^2$ such that
the following duality between pullback vector bundles holds:
\begin{equation}\label{eq_GD_04}
 \varphi_{(1,N)}^\ast L_{q,r}(p) \ \simeq\  \varphi_{(1,M_0)}^\ast{L}_\text{{\upshape ref}}(p)\otimes\text{\upshape det}({E}_{N,q}).
\end{equation}
Here the maps $\varphi_{(n)}:\num{T}^2\to\num{T}^2$, with $n:=(n_1,n_2)\in\Z$ are defined by
\begin{equation}\label{eq_appBBB}
\varphi_{(n)}(\expo{\ii 2\pi k_1},\expo{\ii 2\pi k_2}) = (\expo{\ii 2\pi n_1 k_1},\expo{\ii 2\pi n_2 k_2}),
\end{equation}
$M_0:=qM-rN$, the symbol $\simeq$ denotes isomorphism of vector bundles over $\num{T}^2$ and $\text{\upshape det}({E}_{N,q})\to\num{T}^2$ is the \emph{determinat line bundle} of the vector bundle ${E}_{N,q}$.
\end{teo}

The definition of the vector bundle ${L}_\text{{\upshape ref}}(p)$ and the proof of the geometric duality \eqref{eq_GD_04} are in Sect.~\ref{sec_geoduality}.
For now, out of \eqref{eq_GD_04} we read, for the corresponding characteristic classes,
\begin{equation}\label{sec2_eq_compt}
 c_1(\varphi_{(1,N)}^\ast L_{q,r}(p))=c_1( \varphi_{(1,M_0)}^\ast {L}_\text{{\upshape ref}}(p))+
 \text{Rk}({L}_\text{{\upshape ref}}(p)) \ c_1(\text{\upshape det}({E}_{N,q})).
\end{equation}
Equation \eqref{sec2_eq_compt} follows from the product formula $ch(E_1\otimes E_2)=ch(E_1)\wedge ch(E_2)$ for the Chern character and observing that $ch(E)=\text{Rk}(E)+c_1(E)$ for vector bundles with the base manifold $\num{T}^2$. 
Also, a line bundle has rank 1 while ${L}_\text{{\upshape ref}}(p)$ has the same rank as any of its pull-back. Upon integrating over $\num{T}^2$ we get the corresponding Chern numbers.
Preliminary, we need a classical result in differential geometry: 
\begin{lem}\label{ch5_lem_fuc_chern}
 Let ${E}\to \num{T}^2$ be a Hermitian vector bundle and denote with
$\varphi_{(n)}^\ast {E}\to \num{T}^2$ the pullback with respect to the function \eqref{eq_appBBB}.
Then, the first Chern numbers are related by 
\begin{equation}\label{eq_appBBB1}
C_1(\varphi_{(n)}^\ast {E})=n_1n_2\ C_1( {E}) ,\qquad\qquad  n:=(n_1,n_2)\in \Z^2.
\end{equation}
\end{lem}
\noindent
Using this results to 
$$
C_1(\varphi_{(1,N)}^\ast L_{q,r}(p)) = N\ C_1(L_{q,r}(p)) \qquad \textup{and} \qquad
C_1(\varphi_{(1,M_0)}^\ast{L}_\text{{\upshape ref}}(p))=M_0\ 
C_1({L}_\text{{\upshape ref}}(p))
$$
which, together with the identity $C_1(\text{\upshape det}({E}_{N,q}))=C_1({E}_{N,q})=q$, leads to 
\begin{equation}\label{geom_TKNN}
C_1(L_{q,r}(p)) = q\left[\frac{1}{N}\ \text{Rk}({L}_\text{{\upshape ref}}(p))+\left(\frac{M}{N}-\frac{r}{q}\right) C_1({L}_\text{{\upshape ref}}(p)) \right].
\end{equation}
This geometric equation is at the core of the proof, given in full detail in 
Sect.~\ref{sec_RBR}, of the main Theorem \ref{teo:new1}. As shown there, the key point consists in realizing that the reference bundle ${L}_\text{{\upshape ref}}(p)\to\num{T}^2$ has rank $\text{\upshape Rk}({L}_\text{{\upshape ref}}(p)):=N\ncint(p)$ and (first) Chern number $C_1({L}_\text{{\upshape ref}}(p)) =\ \ncC(p)$, thus translating equation \eqref{geom_TKNN}  to \emph{its natural form}
in equation \eqref{eq:new1}. 

As mentioned in Sect.~\ref{se:intro}, equation \eqref{geom_TKNN} (or equivalently \eqref{eq:new1}) generalizes the TKNN-equations \eqref{eq_010} (in its natural form \eqref{eq_011}). Then,  for operators $H^{\theta}_{q,r}:=\Pi_{q,r}(h_\theta)$, with $h_\theta$ as in \eqref{eq_005} and $\theta=\nicefrac{M}{N}$, one gets exactly the generalized dioaphantine equation \eqref{eq:new2}.

\subsection{Additional comments}\label{sec_int_com_rk}

{\bf Isomorphisms and unitary equivalence.}
Any $(q,r)$-Weyl representations $\Pi_{q,r}$ is faithfull, therefore the family of these representations provides a family of pairwise isomorphic realization of the algebra $\sss{A}_\theta$. However, although isomorphic in an algebraic sense, these realizations are not isomorphic in a Hilbertian sense that is to say, they are not unitarily equivalent. A way to see this comes from looking at the \emph{coupling constant}, a positive number associated to the von Neumann algebra generated in each representation which is invariant under unitary equivalences \cite{rieffel-81b}. In particular, with this technique the non unitary equivalence of 
the representations $\Pi_{1,0}$ and $\Pi_0$ has been showed in \cite{emch-96}. 
On the other hand, the Chern numbers $C_{q,r}(p)$ in \eqref{eq:new1} is seen as describing \emph{topological} properties of the representation $\Pi_{q,r}(\sss{A}_{\nicefrac{M}{N}})$ which are invariant under unitary equivalences. Since $C_{q,r}(p)\neq C_{q',r'}(p)$ for any $p\in\text{Proj}(\sss{A}_{\nicefrac{M}{N}})$, if $(q,r)\neq(q',r')$,
it follows that there is no unitary operator $W:\sss{H}_q\to\sss{H}_{q'}$ such that $\Pi_{q',r'}=W\circ\Pi_{q,r}\circ W^{-1}$.\\  

\noindent{\bf The role of the parameter $r$.} Equation \eqref{eq:new1} can be written as
\begin{equation}\label{eq_10bis}
C_{q,r}({p})=q\, C_{1,0}({p})-r \ncC({p}) =C_{q,0}({p})-r \ncC({p}) .
\end{equation}
We see that for the $0$-twisted representations  the value of $C_{q,0}({p})$ is exactly $q$ times the value of $C_{1,0}({p})$. This is in accordance with Remark~\ref{rk_01} when the representation $\Pi_{q,0}$ was shown to be just $q$ copies of the representation $\Pi_1$. 
On the other hand, when $r\neq0$, the value of the Chern number $C_{q,r}({p})$ differs from $C_{q,0}({p})$ by an extra term: $r$ times the \virg{intrinsic} or \virg{abstarct}
value $\ncC({p})$. Thus the parameter $r$ plays the role of a twist-index.\\

\noindent{\bf Extension to the irrational case.}
Let $h_\theta\in\sss{A}_\theta$ be a selfadjoint operator (not necessarily the Hofstadter operator
in \eqref{eq_005}) and let $p_g$ be the spectral projection
associated to  the gap $g$ by means of the Riesz formula as in \eqref{rf}. 
In the rational case $\theta=\nicefrac{M}{N}$ the integers $C_{q,r}(p_g)$  are well defined quantities related to spectral properties of the projection $p_g$. 
One may ask whether these numbers, initially defined for rational values of $\theta$, are stable for small perturbations of the deformation parameter. A possible positive answer goes as follows. For $\theta\in I$, where $I$ is an interval in $\R$, and $(u,v)$ a system of generators for the corresponding algebra $\sss{A}_\theta$, consider a family of selfadjoint elements $h_\theta:=f(u,v)$ with $f\in C^\infty(\num{T}^2)$ a real smooth functions. The functional expression of $h_\theta$ is fixed and $h_\theta$ depends on $\theta$ only through the fundamental commutation relation  which defines $\sss{A}_\theta$.  Suppose that  
$\varepsilon_g$ is a real number not in the spectrum $\sigma(h_\theta)$ of $h_\theta$ for any $\theta\in I$ and denote by 
$p_g(\theta)\in\text{Proj}(\sss{A}_\theta)$ the related spectral projections for the interval  
$(-\infty,\varepsilon_g]\cap\sigma(h_\theta)$. The functions $\theta\mapsto\ \ncC(p_g(\theta))$
is constant in the interval $I$  \cite[Prop.~11.11]{boca-01}.
On the other hand, from the description of the group $K_0(\sss{A}_\theta)$ given in   \cite{pimsner-voiculescu-80}, one deduces (cf. \cite{connes-94}) that 
\begin{equation}\label{eq_c10b-bis}
\nint\big(p_g(\theta)\big)=m\big( p_g(\theta)\big) - \theta \ncC\big(p_g(\theta)\big) ,
\end{equation}
 where the integer $m(\cdot)\in\Z$ is uniquely determined by the condition
$0\leqslant\ncint(\cdot)\leqslant1$. Equation \eqref{eq_c10b-bis}, implies that the integer $m(\cdot)$ is constant for small perturbation of $\theta$. Hence,
 formula
\begin{equation}\label{eq_11}
C_{q,r}(p_g) = q\left[m(p_g)-\frac{r}{q} \ncC(p_g)\right] = 
q\left[\nint(p_g)+\left(\theta-\frac{r}{q}\right) \ncC(p_g)\right]\in\Z
\end{equation}
is well defined and extends \eqref{eq:new1} also for the irrational values of $\theta\in I$. 
Its interpretation as an equation for conductances for the Hofstadter operator $h_\theta$ in \eqref{eq_005} in the 
$(q,r)$-representation, leads to a generalization of \eqref{eq:new2} as 
\begin{equation}\label{irr-TKNN-eqs}
 t_g+(q \theta - r) s_g = q \, d_g, \qquad\quad g=0,\ldots,N_\text{max}. 
\end{equation}
Now $t_g=C_{q,r}(p_g)$  and $s_g= -\ncC(p_g)$ as before, whereas $d_g =\ \ncint(p_g)$,   
with $p_g$ once again the spectral projections of the Hofstadter operator $h_\theta$. 
\\

\noindent{\bf Cohomological interpretation.}
From a noncommutative geometric point of view, equation \eqref{eq:new1} (and the more general \eqref{eq_11})  are related to the \emph{periodic cyclic homology} of the (smooth) algebra 
$\sss{A}_\theta$. This is the $\Z_2$ graded group
${PH}^\bullet(\sss{A}_\theta):={PH}^\text{ev}(\sss{A}_\theta)\oplus {PH}^\text{od}(\sss{A}_\theta)$ with both component being isomorphic to $\C^2$ \cite{connes-94}. Since two independent generators of ${PH}^\text{ev}(\sss{A}_\theta)$ are the noncommutative trace $\ncint\ $ and the Connes--Chern map  $\ncC$ ,  the integer valued functions $C_{q,r}$ is an even element of the periodic cyclic cohomology group of $\sss{A}_\theta$.

\section{Bloch-Floquet transform and  bundle representation}\label{sec_BF}

In order to get a bundle representation of the $C^\ast$-algebra $\Pi_{r,q}(\sss{A}_{\nicefrac{M}{N}})\subset\bbb{B}(\sss{H}_q)$, thus proving Theorem \ref{teo_main_1}, we use a suitably adapted version \cite{denittis-10,denittis-panati-09} of the Bloch-Floquet theory. The first ingredient for such a theory is a sufficiently rich family of simultaneous symmetries for the Hamiltonians one is considering. Mathematically, this leads to look for  
a maximal-commutative $C^\ast$-subalgebra $\sss{S}^{\nicefrac{M}{N}}_{q,r}$ of the commutant  $\Pi_{r,q}(\sss{A}_{\nicefrac{M}{N}})'$. 
The existence of such an algebra $\sss{S}^{\nicefrac{M}{N}}_{q,r}$ will constructively provide  
the bundle representation for $\Pi_{r,q}(\sss{A}_{\nicefrac{M}{N}})$.

\subsection{The generalized Bloch-Floquet transform} \label{sub_sec_3.1}
As established in \cite{connes-rieffel-87} and in relation with  the previous work \cite{takesaki-69}, for any $\theta\in\R$, the commutant $\Pi_{r,q}(\sss{A}_{\theta})'$ can be identified with a different copy of the NCT-algebra. Since the integers $q$ and $r$ are assumed to be coprime, i.e. \text{g.c.d.}$(q,r)=1$, that there exist a unique  pair of integers  $\alpha,\beta\in\Z$ such that 
\begin{equation}\label{eq_diof1}
\beta q-\alpha r=1,\qquad\qquad \textup{with} \qquad |\alpha|<q .  
\end{equation}
 The commutant $\Pi_{r,q}(\sss{A}_{\theta})'$ is then generated by unitary operators $\widehat{U}^{\theta}_{q,r}$ and $\widehat{V}_q$ acting:
 \begin{equation}\label{eq_007bis}
(\widehat{U}^{\theta}_{q,r}\psi)(x;\ell)=\expo{\ii 2\pi\frac{1}{q}\left(\frac{x}{\epsilon}+{\ell}\alpha\right)}\psi\left(x;\ell\right), \qquad \mathrm{and} \qquad 
(\widehat{V}_q\psi)(x;\ell)=\psi\left(x-\frac{1}{q};[\ell+1]_q\right) .
\end{equation}
One checks that 
\begin{equation}\label{eq_027}
\widehat{U}^{\theta}_{q,r}\ \widehat{V}_q:=\expo{\ii 2\pi{\tilde\theta}}\ \widehat{V}_q\ \widehat{U}^{\theta}_{q,r} \qquad \textup{with} \qquad \tilde\theta(\theta,q,r):=\frac{1}{q^2\epsilon}-\frac{\alpha}{q}=\frac{\beta-\alpha\theta}{q\theta-r}
\end{equation}
and $\alpha=\alpha(q,r)$ and $\beta=\beta(q,r)$ given by \eqref{eq_diof1}.

Using the more compact matrix notation, one also writes
\begin{equation}\label{eq_006bis}
\widehat{U}^{\theta}_{q,r}:=T_1^{\frac{1}{q\epsilon}}\otimes\num{U}^{\alpha}_q,\qquad\qquad\widehat{V}_q=T_2^{\frac{1}{q}}\otimes\num{V}_q^{-1},
\end{equation}
where $T_1$ and $T_2$ are defined by \eqref{eq_004}, and the matrices $\num{U}_q$ and $\num{V}_q$ by \eqref{eq_005bis}.
A straightforward computation shows that the operators $\widehat{U}^{\theta}_{q,r}$ and $\widehat{V}_q$ defined by \eqref{eq_006bis} commute with the operators $U_q$ and $V^\theta_{q,r}$ defined by \eqref{eq_006}. Moreover, in view of the general result \cite[Thm.~3]{takesaki-69}, $\widehat{U}^{\theta}_{q,r}$ and $\widehat{V}_q$ generate by linear combinations and strong topology closure the commutant $\Pi_{r,q}(\sss{A}_{\theta})'$. 
It is worth stressing that the operator $\widehat{U}^{\theta}_{q,r}$ (and then the algebra $\Pi_{r,q}(\sss{A}_{\theta})'$) depends only on $r$ and $q$. Indeed, although the Diophantine equation  \eqref{eq_diof1} (without the restriction $|\alpha|<q$) admits a family of solutions $\alpha_n=\alpha+nq$ and $\beta_n=\beta+nr$ (with $\alpha$ and $\beta$ fixed by the constraint), such an ambiguity is irrelevant in the definition of $\num{U}^{\alpha}_q=\num{U}^{\alpha_n}_q$. Moreover, 
$\frac{\beta_n-\alpha_n\theta}{q\theta-r}=\frac{\beta-\alpha\theta}{q\theta-r}-n$, yielding isomorphic algebras for the commutant. 
Notice that for $q=1$ (which implies $r=0$), one has $\tilde\theta=\theta^{-1}$ and so
$\widehat{U}^{\theta}_{1,0}$ and $\widehat{V}_1$ provide a (faithful) representation of the NCT-algebra $\sss{A}_{\nicefrac{1}{\theta}}$ on the Hilbert space $\sss{H}_1$.

\goodbreak
\medskip

For the rational case, $\theta=\nicefrac{M}{N}$, the deformation parameter in \eqref{eq_027}  becomes
\begin{equation}\label{eq_028}
\tilde\theta=\frac{bN-aM}{M_0}\qquad\text{with}\qquad M_0:=qM-rN\in\Z.
\end{equation}
From equation \eqref{eq_027} it then readily follows that 
\begin{equation}\label{eq_028''}
\widehat{U}^{\nicefrac{M}{N}}_{q,r}\, (\widehat{V}_q )^{M_0} = (\widehat{V}_q )^{M_0} \, \widehat{U}^{\nicefrac{M}{N}}_{q,r} .
\end{equation}
Let  $\sss{S}^{\nicefrac{M}{N}}_{q,r}:=C^\ast(\widehat{U}^{\nicefrac{M}{N}}_{q,r},(\widehat{V}_q )^{M_0})$ be the $C^\ast$-algebra of operators generated by $\widehat{U}^{\nicefrac{M}{N}}_{q,r}$ and $(\widehat{V}_q )^{M_0}$. By construction $\sss{S}_{q,r}^{\nicefrac{M}{N}}$ is a commutative subalgebra of the commutant $\Pi_{q,r}(\sss{A}_{\nicefrac{M}{N}})'$.
It is a straightforward computation to show that $\sss{S}_{q,r}^{\nicefrac{M}{N}}$ is  maximal commutative inside the commutant $\Pi_{q,r}(\sss{A}_{\nicefrac{M}{N}})'$, i.e. it is not properly contained in any other commutative subalgebra of the commutant. We refer to the triple $\{\sss{H}_q,\Pi_{q,r}(\sss{A}_{\nicefrac{M}{N}}),\sss{S}_{q,r}^{\nicefrac{M}{N}}\}$ as the \emph{standard irreducible triple} for the $(q,r)$-Weyl representation of the NCT-algebra $\sss{A}_{\nicefrac{M}{N}}$. 

\begin{rk}\label{rk_101} Using the fact that $\num{V}_q^{-qM}=\num{I}_q$, one also write
\begin{equation}\label{eq_028ter}
(\widehat{V}_q )^{M_0}=T_2^{\frac{M_0}{q}}\otimes\num{V}_q^{rN} . 
\end{equation}
If $N=qN'$ with $N'\in\Z$ and $\text{{\upshape g.c.d.}}(N',q)=1$ then $M_0=qM_0':=q(M-rN')$ and from equation
\eqref{eq_028ter} it also follows that 
\begin{equation}\label{eq_029''}
(\widehat{V}_q )^{M_0}=(\widehat{V}_q)^{qM_0}=T_2^{M_0'}\otimes\num{I}_q.
\end{equation}
In this case both  $(\widehat{V}_q )^{M_0}$ and $\widehat{U}^{\nicefrac{M}{N}}_{q,r}$ are diagonal  with respect to the discrete variable $\ell$ of the space $\sss{H}_q$ and $\sss{S}_{q,r}^{\nicefrac{M}{N}}$
decomposes as a direct sum of $q$ copies of a commutative $C^\ast$-algebra on $\sss{H}_1=L^2(\R,dx)$. To avoid this degenerate situation, we assume henceforth that  $\text{{\upshape g.c.d.}}(N,q)=1$. Since $\text{{\upshape g.c.d.}}(q,r)=1$ it also holds that $\text{{\upshape g.c.d.}}(q,rN)=1$.
\hfill $\blacklozenge\lozenge$
\end{rk}

The $C^\ast$-algebra $\sss{S}_{q,r}^{\nicefrac{M}{N}}$ is isomorphic to the $C^\ast$-algebra $C(\num{T}^2)$ of continuous functions on the two-dimensional torus $\num{T}^2$. 
Preliminarily, with $(m_1,m_2)\in\Z^2$ we conside the unitary operators 
\begin{equation}\label{eq_not}
 {\hat W}_{(m_1,m_2)}={\hat W}_{(m_1,m_2)}(q,r,\nicefrac{M}{N}):=\big( (\widehat{V})^{M_0}_q \big)^{m_1} \big( \widehat{U}^{\nicefrac{M}{N}}_{q,r} \big)^{m_2}=\big( \widehat{U}^{\nicefrac{M}{N}}_{q,r} \big)^{m_2}\big( (\widehat{V})^{M_0}_q \big)^{m_1},
\end{equation}
whose explicit action is 
\begin{align}\label{future-use}
\left({\hat W}_{(m_1,m_2)}\psi\right)(x;\ell)&=\expo{\ii 2\pi\left(\frac{N}{M_0} x+\frac{\alpha}{q} \ell \right) m_2} \, \psi 
\Big(x-\frac{M_0}{q}m_1 ; [\ell+M_0 m_1]_q \Big) \nonumber \\ 
&=\expo{\ii 2\pi\left(\frac{N}{M_0} x+\frac{\alpha}{q} \ell \right) m_2} \, \psi 
\Big(x-\frac{M_0}{q}m_1 ; [\ell - r N m_1]_q \Big).
\end{align}
{
The operators ${\hat W}_{(m_1,m_2)}$ generates the commutative $C^\ast$-algebra $\sss{S}_{q,r}^{\nicefrac{M}{N}}$. More precisely, the following holds \cite[Prop.~4.5.7]{denittis-10}. 
\begin{propos}
The operators  ${\hat W}_{(m_1,m_2)}$ yield a faithful unitary representation of $\Z^2$ on $\sss{H}_q$ such that 
$$
\sum_{(m_1,m_2)\in\Z^2} c_{(m_1,m_2)}\ {\hat W}_{(m_1,m_2)}=0\qquad\Leftrightarrow\qquad\ c_{(m_1,m_2)}=0, \quad \forall\, (m_1,m_2)\in\Z^2.
$$ 
Moreover the Gel'fand spectrum of $\sss{S}_{q,r}^{\nicefrac{M}{N}}$ coincides with the two-dimensional torus $\num{T}^2$ and the \emph{basic} measure coincides with the (normalized) Haar measure $\dd z:=\dd k_1\wedge \dd k_2$.
\end{propos}
}
The proof of the above result is based on what, in the language of \cite{denittis-10,denittis-panati-09} (and more generally in the theory of shift operators \cite{nagy-foias-10}), is called a \emph{wandering system} of cardinality $N$ for the commutative $C^\ast$-algebra $\sss{S}_{q,r}^{\nicefrac{M}{N}}$, that we are about to describe. 
For this, we need to cover $ \R$ with intervals $I_{j,n}\subset\R$ defined, for any $j,n\in\Z$, by
\begin{equation}\label{eq_wand_har_bis_bos}
I_{j,n}:=\left\{
\begin{aligned}
&\left[j\frac{M_0}{N}+nM_0;(j+1)\frac{M_0}{N}+nM_0\right)&&\ \ \text{if}\ \ \ M_0>0\\
\\
&\left((j+1)\frac{M_0}{N}+nM_0;j\frac{M_0}{N}+nM_0\right]&&\ \ \text{if}\ \ \ M_0<0.
\end{aligned}
\right. 
\end{equation}
Let $\{\Upsilon_0,\ldots,\Upsilon_{N-1}\}\subset\sss{H}_q$ be the set of $N$ vectors defined  by
\begin{equation}\label{eq_wand_vec_01}
\Upsilon_j(\cdot)=\left(
\begin{array}{c}
\gamma_j(\cdot\ ;0) \\ 
\gamma_j(\cdot\ ;1) \\ 
\vdots \\ 
\gamma_j(\cdot\ ;q-1)                        \end{array}\right):=\left(
\begin{array}{c}
\chi_{j,0}(\cdot) \\ 
 0 \\ 
\vdots \\ 
0                        \end{array}\right),\qquad\qquad j=0,\ldots,N-1.
\end{equation}
with characteristic functions
\begin{equation}\label{eq_wand_har}
\chi_{j,n}(x):=\left\{
\begin{aligned}
&\sqrt{\frac{N}{|M_0|}}&&\ \ \text{if}\ \ \ x\in I_{j,n}\\
\\
&0&&\ \ \text{otherwise} .
\end{aligned}
\right. 
\end{equation}

\noindent
From the definition it follows that $(\Upsilon_i;\Upsilon_j)_{\sss{H}_q}=\delta_{i,j}$. Moreover,  $q$ and $rN$ being coprime, $(\widehat{V}^{M_0}_q)^m\Upsilon_i$ and  $(\widehat{V}^{M_0}_q)^{m'}\Upsilon_j$ have  the non-null component labeled by the same index  if and only if $m-m'\in q\Z$; also $(\widehat{V}^{M_0}_q)^m\Upsilon_i$ and  $(\widehat{V}^{M_0}_q)^{m+pq}\Upsilon_j$ are supported on disjoint sets whenever $p\in\Z\setminus\{0\}$.
On the other hand, each interval $I_{j,n}$ has length $\nicefrac{|M_0|}{N}=|q\epsilon|$. If $I$ is any such a subinterval, the family of functions $\expo{\ii 2\pi\frac{N}{M_0}mx}\chi_{I}$, with $m\in\Z$, 
and $\chi_{I}$ the (normalized) characteristic function of $I$,
provides an orthonormal basis for $L^2(I)$. By using all of this, and \eqref{future-use},  it follows that 
\begin{equation}\label{eq_wand_vec_02}
(\Upsilon_i;{\hat W}_{(m_1,m_2)}\Upsilon_j)_{\sss{H}_q}=\delta_{i,j}\ 
\delta_{m_1,0}\ \delta_{m_2,0}, \qquad\qquad 0\leqslant i,j\leqslant N-1,\qquad (m_1,m_2)\in\Z^2
\end{equation}
for ${\hat W}_{(m_1,m_2)}$ the unitary operators in \eqref{eq_not}. Furthermore, using the obvious identification
\begin{equation}\label{ch5_int_dec_har}
\R=\bigcup_{j=0}^{N-1}\bigcup_{n\in\Z}I_{j,n},\qquad\ \Rightarrow\ \qquad L^2(\R)\simeq\bigoplus_{j=0}^{N-1}\bigoplus_{n\in\Z}L^2(I_{j,n})
\end{equation}
simple computations show that the space
\begin{equation}\label{eq_wand_vec_03}
\Phi_{q,r} := \left\{{\hat W}_{(m_1,m_2)}\Upsilon_j\ :\  j=0,\ldots, N-1,\qquad (m_1,m_2)\in\Z^2\right\}
\end{equation}
is dense in $\sss{H}_q$.
As mentioned, in the language of the theory of  shift operators, properties \eqref{eq_wand_vec_02} and \eqref{eq_wand_vec_03} are what makes the set of orthonormal vectors
$\{\Upsilon_0,\ldots,\Upsilon_{N-1}\}$ a wandering system of cardinality $N$ for the commutative $C^\ast$-algebra $\sss{S}_{q,r}^{\nicefrac{M}{N}}$.

The space $\Phi_{q,r}$ can be realized as  the inductive limit of finite dimensional Hilbert spaces
$\sss{H}_{q,r}(m)\subset\sss{H}_q$ defined as in \eqref{eq_wand_vec_03} for any $m\in\N$ with restriction 
$0\leqslant |m_1|+|m_2|\leqslant m$.  Clearly $\sss{H}_{q,r}(m)\subset\sss{H}_{q,r}(m+1)$  and $\sss{H}_{q,r}(0)$ coincides with the linear span of the wandering vectors $\{\Upsilon_0,\ldots,\Upsilon_{N-1}\}$ (hence it is $N$ dimensional). As a set $\Phi_{q,r}\subset\sss{H}_q$ is made of finite linear combinations of the orthonormal basis generated by the action of the unitaries ${\hat W}_{(m_1,m_2)}$ on vectors of the wandering system. Once $\Phi_{q,r}$ is endowed with the strict inductive limit topology, it becomes a nuclear space dense in $\sss{H}_q$ with respect to the norm topology and the embedding $\jmath :\Phi_{q,r}\hookrightarrow\sss{H}_q$ is norm continuous.

Next, let $\Phi_{q,r}^\ast$ be the topological dual of $\Phi_{q,r}$ endowed with the $\ast$-weak topology. Clearly, $\Phi_{q,r}^\ast$ is a space of distributions with $\Phi_{q,r}\hookrightarrow\sss{H}_q=\sss{H}_q^\ast\hookrightarrow \Phi_{q,r}^\ast$ and the dual pairing between $\Phi_{q,r}$ and $\Phi_{q,r}^\ast$ is compatible with the scalar product of $\sss{H}_q$. A \emph{generalized Bloch-Floquet transform}  \cite[Sect.~4.6]{denittis-10} between $\Phi_{q,r}$ and $\Phi_{q,r}^\ast$ is given by defining for any $k:=(k_1,k_2)\in \R^2$ a linear map $\left.\bbb{F}_{q,r}\right|_{k}:\Phi_{q,r}\to\Phi_{q,r}^\ast$ by the formula
\begin{equation}\label{eq_031}
\left.\bbb{F}_{q,r}\right|_{k}:\Xi\longmapsto(\bbb{F}_{q,r}\Xi)(k):=\sum_{(m_1,m_2)\in\Z^2}\expo{- \ii 2\pi k_1m_1}
\expo{- \ii 2\pi k_2m_2}\ {\hat W}_{(m_1,m_2)}\ \Xi 
\end{equation}
where the operators ${\hat W}_{(m_1,m_2)}$ are defined in \eqref{eq_not}.
Denote with $\zeta^{j}_{(q,r)}(k):=(\bbb{F}_{q,r}\Upsilon_j)(k)$ the Bloch-Floquet transform of the $j$-th wandering vector $\Upsilon_j$.  The collection 
\begin{equation}\label{eq_frame}
\boldsymbol{\zeta}_{(q,r)}(k):=\left\{ \zeta^{0}_{(q,r)}(k),\ldots,\zeta^{N-1}_{(q,r)}(k) \right\}\subset \Phi_{q,r}^\ast \qquad \textup{for any} \quad k:=(k_1,k_2) \in \R^2 , 
\end{equation}
yields a frame of $N$ linear independent distributions which spans an $N$-dimensional complex vector space $\sss{H}_{q,r}(k)\subset \Phi_{q,r}^\ast$. An Hermitian structure on $\sss{H}_{q,r}(k)$ is defined by assuming $\boldsymbol{\zeta}_{(q,r)}(k)$ to be an orthonormal frame, namely 
$(\zeta^{i}_{(q,r)}(k);\zeta^{j}_{(q,r)}(k))_k:=\delta_{i,j}$.

As we shall show below, the $N$-dimensional complex vector spaces $\sss{H}_{q,r}(k)\subset \Phi_{q,r}^\ast$ obey  (pseudo-)periodic conditions allowing one to glue them together to a bundle over the torus $\num{T}^2$. For the moment, we mention that from its very definition, the \virg{fiber spaces} $\sss{H}_{q,r}(k)$ are simultaneous (distributional) eigenspaces for the element of  the commutative algebra $\sss{S}_{q,r}^{\nicefrac{M}{N}}$. This means that by using the Gel'fand isomorphism  $\sss{S}_{q,r}^{\nicefrac{M}{N}}\ni A_f\leftrightarrow f\in C(\num{T}^2)$ to label elements in $\sss{S}_{q,r}^{\nicefrac{M}{N}}$, one has that
$$
\bbb{F}_{q,r}(A_f\Psi)(k)=f(\expo{\ii 2\pi k_1}, \expo{\ii 2\pi k_2}) (\bbb{F}_{q,r}\Psi)(k) , \qquad\qquad \textup{for} 
\qquad \Psi\in\sss{H}_q
$$

We proceed by explicitly computing the functional form of the frame   $\boldsymbol{\zeta}_{(q,r)}(k)$.
\begin{propos}\label{prop:frame}
For any $k=(k_1,k_2)\in\R^2$, the action of the distributions 
$$
\zeta^j_{(q,r)}(k):=\left(
\begin{array}{c}
\zeta^j_{(q,r)}(k)|(\cdot\ ;0) \\ 
\zeta^j_{(q,r)}(k)|(\cdot\ ;1)\\
\vdots \\ 
\zeta^j_{(q,r)}(k)|(\cdot\ ;q-1)
\end{array}
\right)\in \Phi_{q,r}^\ast\qquad\qquad  j=0,\ldots,N-1
$$ 
 on  $\Phi_{q,r}$ is given by
 \begin{equation}\label{eq_sec_har_01bis}
\zeta^j_{(q,r)}(k)|(\cdot\ ;\ell):=\sqrt{\frac{|M_0|}{N}}\ \sum_{m\in\Z}\expo{- \ii2\pi k_1(\tau_\ell+mq)}\ \delta\left[\cdot\ -\frac{M_0}{N}(k_2+j) -mM_0-\tau_\ell\frac{M_0}{q}\right]
\end{equation} 
where for $\ell\in\{0,\ldots,q-1\}$, the permutation $\tau:\ell\mapsto\tau_\ell$ is defined  by $\ell=[\tau_\ell rN]_q$
and the Dirac delta function $\delta(\cdot-x_0)$ acts on functions $\xi:\R\to\C$ as the evaluation in $x_0$, i.e. $\langle \delta(\cdot-x_0);\xi\rangle=\xi(x_0)$.
\end{propos}
\Proof
Let $\Upsilon_{j}(\cdot)$ be given by \eqref{eq_wand_vec_01}
and ${\hat W}_{(m,n)}$ by \eqref{eq_not}.
Using the identification  between the dual piring $\langle\cdot;\cdot\rangle:\Phi_{q,r}^\ast\times\Phi_{q,r}\to\C$ and the scalar product on $\sss{H}_q$ for distributions in $\sss{H}_q\cap\Phi_{q,r}^\ast$, for any $\Xi(\cdot):=(\xi(\cdot;0),\ldots,\xi(\cdot;q-1))\in\Phi_{q,r}$, one has
\begin{align}\label{eq_p01}
\langle{\hat W}_{(m,n)}\ \Upsilon_{j};\Xi\rangle:&=\left(\Upsilon_{j};{\hat W}_{(-m,-n)}\  \Xi\right)_{\sss{H}_q}=\sum_{\ell=0}^{q-1}\left(\gamma_j(\cdot;\ell);\left({\hat W}_{(-m,-n)}\xi\right)(\cdot;\ell)\right)_{L^2(\R)}\\
&=\left(\chi_{j,0}(\cdot);\left({\hat W}_{(-m,-n)}\xi\right)(\cdot;0)\right)_{L^2(\R)} . \nonumber
\end{align}
From \eqref{future-use} one gets
\begin{equation}\label{eq_p02}
\left({\hat W}_{(-m,-n)}\xi\right)(x;0)= \expo{-\ii2\pi \frac{N}{M_0} \, nx}\ \xi\left(x+m\frac{M_0}{q}; [mrN]_q\right) 
\end{equation}
and in turn \eqref{eq_p01} becomes 
\begin{align}\label{eq_p001'}
\langle{\hat W}_{(m,n)}\ \Upsilon_{j};\Xi\rangle&:=
\int_{\R}\chi_{j,0}(x)\  \expo{-\ii2\pi \frac{N}{M_0} \, nx}\ \xi\left(x+m\frac{M_0}{q}; [mrN]_q\right) \ \dd x \\
 \nonumber \\  
&\: =  \sqrt{\frac{N}{|M_0|}} \int_{I_{j,0}} \expo{-\ii2\pi \frac{N}{M_0} \, nx}\ \xi\left(x+m\frac{M_0}{q}; [mrN]_q\right) \ \dd x \, . \nonumber \\  
\nonumber \\  
&\: =  \sqrt{\frac{N}{|M_0|}} \int_{I_{0,0}} \expo{-\ii2\pi \frac{N}{M_0} \, nx}\ \xi\left(x+ j \frac{M_0}{N} + m\frac{M_0}{q}; [mrN]_q\right) \ \dd x \, . \nonumber
\end{align}
with the domains $I_{j,0}$ and $I_{0,0}$ defined by \eqref{eq_wand_har_bis_bos}.
This integral is just proportional to the $n$-th Fourier coefficient of the function under the integral. In fact, from the very definition of the space $\Phi_{q,r}$ the $[mrN]_q$-th component of $\Xi$ is a trigonometric polynomial when restricted to any shifted intervals $I_{j,n}+m\frac{M_0}{q}$. More precisely, 
$$
\langle{\hat W}_{(m,n)}\ \Upsilon_{j};\Xi\rangle= \sqrt{\frac{{|M_0|}}{N}} \ \hat\phi_{n}
$$
with 
\begin{equation}\label{eq_p004}
\hat\phi_n:= \frac{N}{|M_0|} \int_{I_{0,0}} \expo{-\ii2\pi \frac{N}{M_0} \, nx}\ \xi\left(x+ j \frac{M_0}{N} + m\frac{M_0}{q}; [mrN]_q\right) \ \dd x \, .
\end{equation}
Then
\begin{align*}
\langle\sum_{n\in\Z}
\expo{- \ii2\pi k_2 n }{\hat W}_{(m,n)}\ \Upsilon_{j};\Xi\rangle
=\sqrt{\frac{|M_0|}{N}} \, \sum_{n\in\Z} \expo{ \ii2\pi k_2 n}\ \hat\phi_{n}   
=\sqrt{\frac{|M_0|}{N}} \, \sum_{n\in\Z} \hat\phi_{n}\  \expo{ \ii2\pi \frac{N}{M_0} \, n \, \left(\frac{M_0}{N} k_2 \right)}
\end{align*}
which is just the relevant Fourier expansion. Thus, one obtains 
\begin{equation}\label{eq_aperiod}
\langle\sum_{n\in\Z}
\expo{-\ii 2\pi k_2n}{\hat W}_{(n,m)}\ \Upsilon_{j};\Xi\rangle=\sqrt{\frac{|M_0|}{N}}\ 
\xi\left(\frac{M_0}{N}(k_2+j)+m\frac{M_0}{q};[mrN]_q\right) .
\end{equation}
There is an apparent mismatch between the left-hand side of \eqref{eq_aperiod} being formally periodic in $k_2$ of the  versus the corresponding non-periodicity of the right-hand side. This is compensate by a corresponding shift in the 
$\Upsilon_{j}$'s. The translation $k_2\mapsto k_2+1$ results in the distributional shift $\Upsilon_{j}\mapsto \Upsilon_{j+1}$ and so on.

Since $\zeta^j_{(q,r)}:=\bbb{F}_{q,r}\Upsilon_j$ with the generalized Bloch-Floquet transform in \eqref{eq_031}, it follows
\begin{align}\label{eq_smart1}
\langle\zeta^j_{(q,r)}(k);\Xi\rangle = \sqrt{\frac{|M_0|}{N}}\ \sum_{m\in\Z}\expo{\ii2\pi k_1m}\ 
\xi\left(\frac{M_0}{N}(k_2+j)+m\frac{M_0}{q};[mrN]_q\right) .
\end{align}
If $\sigma_m\in\{0,\ldots,q-1\}$ is the representative of the class $[mrN]_q$ (recall that $q$ and $rN$ are coprime), upon shifting $m \mapsto \ell + mq$, this can be written as
\begin{align}\label{eq_smart2}
\langle\zeta^j_{(q,r)}(k);\Xi\rangle=\sum_{\ell=0}^{q-1}\ \sqrt{\frac{|M_0|}{N}}\ \sum_{m\in\Z}\expo{\ii2\pi k_1(\ell+mq)}\ \xi\left(\frac{M_0}{N}(k_2+j)+m M_0 + \ell\frac{M_0}{q};\sigma_\ell\right)
\end{align}
being $\sigma_{(\ell+mq)}=\sigma_\ell$ by definition.
The map $\sigma:\ell\mapsto\sigma_\ell:=[\ell r N]_q$ defines a permutation of the set $\{0,1,\ldots,q-1\}$. If $\tau$ 
is the inverse permutation, $\sigma_{\tau_\ell}=\tau_{\sigma_\ell}=\ell$, we have finally
\begin{align*}
\langle\zeta^j_{(q,r)}(k);\Xi\rangle=\sum_{\ell=0}^{q-1}\ \sqrt{\frac{|M_0|}{N}}\ \sum_{m\in\Z}\expo{ \ii2\pi k_1(\tau_\ell+mq)}\ \xi\left(\frac{M_0}{N}(k_2+j)+m M_0 + \tau_\ell\frac{M_0}{q}; \ell\right)
\end{align*}
which implies \eqref{eq_sec_har_01bis}.\CVD

\goodbreak
\medskip
\begin{propos}\label{ppc}
Let $\boldsymbol{\zeta}_{(q,r)}(\cdot):=\{{\zeta}_{(q,r)}^0(\cdot),\ldots,{\zeta}_{(q,r)}^{N-1}(\cdot)\}$. 
For any $n=(n_1,n_2)\in\Z^2$, these maps obey the pseudo-periodic conditions
\begin{equation}\label{eq_012}
\boldsymbol{\zeta}_{(q,r)}(k+n)= \left( \num{G}_{N,q}(k) \right)^{n_2}\cdot \boldsymbol{\zeta}_{(q,r)}(k),
\end{equation}
where the $N\times N$ unitary matrix $\num{G}_{N,q}(k)$ is given by
\begin{equation}\label{eq_013}
\num{G}_{N,q}(k):=
\left(
\begin{array}{cccl}
0 &  1 & \ldots  &  0 \\ 
0 & 0 &  \ddots  & \vdots \\ 
 \vdots & \vdots  & \ddots  & 1 \\ 
 \expo{ \ii2\pi qk_1} &  \ldots  & 0 & 0
\end{array}
\right)
\end{equation}
\end{propos}
\Proof
From the expression \eqref{eq_sec_har_01bis}, it is clear that each $\zeta^j_{(q,r)}(\cdot)$ is left invariant upon translating the first variable $k_1 \mapsto k_1 + n_1$. On the other hand, for the second variable, 
$$
\zeta^j_{(q,r)}(k_1,k_2 + n_2)=
\left\{
\begin{aligned}
&\expo{ \ii2\pi qk_1\, s } \, \zeta^{j+\tilde{n}_2}_{(q,r)}(k_1,k_2) &&\ \ \text{if}\ \ \ j=0, \cdots, N-1-\tilde{n}_2\\
\\
&\expo{ \ii2\pi qk_1\, (s+1)} \, \zeta^{j+\tilde{n}_2-N}_{(q,r)}(k_1,k_2) &&\ \ \text{if}\ \ \ j=N-\tilde{n}_2, \cdots, N -1 ,
\end{aligned}
\right. 
$$
where $\tilde{n}_2 \in \{ 0,1, \cdots, N-1 \}$, and $n_2 = \tilde{n}_2 + N s$ for some integer $s$. Simple manipulations lead to the transformation \eqref{eq_012} for the vector $\boldsymbol{\zeta}_{(q,r)}(\cdot)$, with the matrix in \eqref{eq_013}.
\CVD


\subsection{The ambient vector bundle}\label{sec_loc_triv}
A immediate consequence  of Proposition \ref{ppc} is that  the frame 
$\boldsymbol{\zeta}_{(q,r)}(k)$ with the pseudo-periodic conditions \eqref{eq_012} are enough to describe the geometric structure of the vector bundle ${E}_{N,q}\to\num{T}^2$. First of all,
conditions \eqref{eq_012} imply that $\sss{H}_{q,r}(k+n)=\sss{H}_{q,r}(k)$ for any $k\in\R^2$ and $n\in\Z^2$, i.e. the $N$-dimensional complex vector space $\sss{H}_{q,r}(k)$ depends only on the equivalence class $[k]\in\R^2/ \Z^2$. 
Let us consider now the projection
\begin{equation}\label{proj}
P_{N,q}(k):=\sum_{j=0}^{N-1}\ketbra{\zeta^j_{(q,r)}(k)}{\zeta^j_{(q,r)}(k)} .
\end{equation}
From Proposition~\ref{ppc}, for any $n=(n_1,n_2)\in\Z^2$, this projection transforms as
\begin{equation}\label{proj-bis}
P_{N,q}(k+n)= \left( \num{G}_{N,q}(k) \right)^{n_2} P_{N,q}(k) \left( \num{G}_{N,q}(k) \right)^{-n_2}.
\end{equation}
As a consequence, the K-theory class of $P_{N,q}(k)$ again depends only on the equivalence class $[k]\in\R^2/ \Z^2$, that is to say it describes a vector bundle over the torus, $\iota: {E}_{N,q}\to\num{T}^2$, with the fibers of the bundles just given by
\begin{equation}\label{fib-can}
\iota^{-1}(k) =P_{N,q}(k) \big( \Phi^\ast_{q,r} \big) \simeq \sss{H}_{q,r}(k). 
\end{equation}
The projection $P_{N,q}(k)$ also provides a canonical covariant derivative on the bundle ${E}_{N,q}$ (named in a variety of ways:  Berry, Grassmannian, Levi-Civita, ...), given by $\nabla^{(\text{B})}  = P \circ \dd$. The corresponding connection 1-form has components:
\begin{equation}\label{eq_berry_con}
\omega_{i,j}^{(\text{B})}(k):=\bra{\zeta_{(q,r)}^i(k)}\ \nabla^{(N,q)}\ \ket{ \zeta_{(q,r)}^j(k)}=
 \braket{\zeta_{(q,r)}^i(k)}{\dd \zeta_{(q,r)}^j(k)}\ ,
\end{equation}
for $i,j=0,\ldots,N-1$. The related  curvature  
$(\nabla^{(\text{B})})^2=P_{N,q}(\dd P_{N,q}\wedge \dd P_{N,q})$, 
can be used to compute the first Chern number of the bundles as 
$$
C_1(E_{N,q})= \frac{\ii}{2\pi} \int_{\num{T}^2} \text{Tr}_N[ (\nabla^{(\text{B})})^2 ]
$$
with 
\begin{equation}\label{eq_berry_conbis}
\text{Tr}_N[(\nabla^{(\text{B})})^2 ](k):=\sum_{j=0}^{N-1}\bra{\zeta_{(q,r)}^j(k)}\ P_{N,q}(\dd P_{N,q}\wedge \dd P_{N,q})\ \ket{ \zeta_{(q,r)}^j(k)}.
\end{equation}
To avoid the need to differentiate delta functions in a computation of the connection \eqref{eq_berry_con}, we use a simpler curvature by means of charts. This way, albeit less elegant, uses explicitly our knowledge of the geometry of the vector bundle ${E}_{N,q}$. The computation is relegated in App.~\ref{se:ccn} and leads to the stated result, i.e.
$$
C_1(E_{N,q}) = q. 
$$

\subsection{Fibered and bundle representations}\label{sec_bund_rep1}

The direct integral $\int^\oplus_{\num{T}^2} \sss{H}_{q,r}(k)\ \dd z(k)$ of the spaces $\sss{H}_{q,r}(k)$, is the collection of \emph{vector fields} $\varphi(\cdot)$ associating to any $k\in \num{T}^2$ a vector $\varphi(k)\in \sss{H}_{q,r}(k)$, such that
\begin{equation}\label{eq_032}
\|\varphi(\cdot)\|^2:=\int_{\num{T}^2}(\varphi(k);\varphi(k))_k\  {\dd k_1\wedge \dd k_2}<+\infty.
\end{equation}
The norm \eqref{eq_032} (and the related scalar product) endows $\int^\oplus_{\num{T}^2} \sss{H}_{q,r}(k)\ \dd z(k)$ with  a Hilbert space structure. Moreover, since the frame $\boldsymbol{\zeta}_{(q,r)}(\cdot)$ is also the basis of this direct integral decomposition, we have the following (tautological) identification:
\begin{equation}\label{eq_t001}
 \int^\oplus_{\num{T}^2} \sss{H}_{q,r}(k)\ \dd z(k)\cong L^2({E}_{N,q})
\end{equation} 
where $L^2({E}_{N,q})$ denotes the set of the \emph{$L^2$-section} of the vector bundle ${E}_{N,q}$.

The generalized Bloch-Floquet transform \eqref{eq_031} extends \cite[Thms.~4.6.2 and 4.6.4]{denittis-10}
to a unitary isomorphism of Hilbert spaces
\begin{equation}\label{eq_0320}
 \bbb{F}_{q,r}:\sss{H}_q\longrightarrow L^2({E}_{N,q}).
\end{equation}
Also, the  transform \eqref{eq_031} allows one to to define a family of irreducible representations 
$\pi_k:\sss{A}_{\nicefrac{M}{N}}\to\bbb{B}(\sss{H}_{q,r}(k))\simeq\text{Mat}_N(\C)$ of the NCT-algebra,  labelled by $k\in\num{T}^2$, and 
glued together via the direct integral structure to a faithfull \emph{fibered representation} of $\sss{A}_{\nicefrac{M}{N}}$, 
\begin{equation}
\widetilde\Pi_{q,r}: \sss{A}_{\nicefrac{M}{N}} \longrightarrow \bbb{B}\left(L^2(E_{N,q})\right), 
\qquad a \longmapsto \int^\oplus_{\num{T}^2} \pi_k(a)\ \dd z(k) .
\end{equation}
In fact, one has
$\widetilde\Pi_{q,r}(\sss{A}_{\nicefrac{M}{N}})\subset\text{End}_{C(\num{T}^2)}(\Gamma(E_{N,q}))\simeq \Gamma(\text{End}(E_{N,q}))$ (as needed for Theorem~\ref{teo_main_1}), so that $\widetilde\Pi_{q,r}$ is a \emph{bundle representation}. 

Each representation $\pi_k$ is defined by
$\pi_k(\cdot)=\left.\bbb{F}_{q,r}\right|_{k}\ \Pi_{q,r}(\cdot)\  {\left.\bbb{F}_{q,r}\right|_{k}}^{-1}$ or, equivalently, 
\begin{equation}\label{eq_p011}
\langle \pi_k(a) \left.\bbb{F}_{q,r}\right|_{k}(\Xi_1);\Xi_2 \rangle=\langle\left.\bbb{F}_{q,r}\right|_{k}(\Xi_1);\Pi_{q,r}(a)^{*}\, \Xi_2 \rangle,\qquad\qquad \textup{with} \quad \Xi_1,\Xi_2\in\Phi_{q,r}.
\end{equation}
for $a\in\sss{A}_{\nicefrac{M}{N}}$. 
 On the basis $\{ \zeta^{j}_{(q,r)}(k) \}$ of $\sss{H}_{q,r}(k)$, this reduces to
\begin{equation}\label{eq_p012}
\langle\pi_{k}(a)\zeta^j_{(q,r)}(k);\Xi \rangle=\langle\zeta^j_{(q,r)}(k);\Pi_{q,r}(a)^{*}\, \Xi \rangle, \qquad\qquad \textup{with} \quad \Xi\in\Phi_{q,r}, 
\end{equation}
for which we may use the explicit formula \eqref{eq_sec_har_01bis}. Owing to the universal property of the NCT-algebra, it is enough to compute this for the generators
\begin{equation}\label{eq_p010}
U_q(k):=\pi_k(u),\qquad\qquad
 V_{q,r}^{\nicefrac{M}{N}}(k):=\pi_k( v) .
\end{equation}
For  $U_q=\Pi_{q,r}(u)$, using its action \eqref{eq_007} in \eqref{eq_p012}, a straightforward computation yields
\begin{equation}\label{eq_p013}
U_q(k)\ \zeta^j_{(q,r)}(k)=\expo{\ii2\pi  \left( \frac{M_0}{N}k_2 + \frac{q M}{N} j  \right)} \zeta^j_{(q,r)}(k), \qquad\qquad j=0,\ldots,N-1.
\end{equation}
Thinking of the $\zeta^j_{(q,r)}(k)$'s as basis vectors, in matrix form:
\begin{equation}\label{matu}
U_q(k) = \expo{\ii2\pi\frac{M_0}{N}k_2 }\ (\num{U}_N)^{qM} ,
\end{equation} 
with the matrix $\num{U}_N$ defined in \eqref{shn}. Clearly $(U_q(k))^N=\expo{\ii2\pi M_0 k_2}\, \num{I}_N$.

For the generator $V_{q,r}^{\nicefrac{M}{N}}=\Pi_{q,r}(v)$, using again its action \eqref{eq_007}, formula \eqref{eq_p012} requires computing the action of the distributions $\zeta^j_{(q,r)}(k)$ on the vectors
\begin{equation}\label{eq_q014}
((V_{q,r}^{\nicefrac{M}{N}})^* \, \Xi)(x;\ell)=\xi\left(x+\frac{1}{q}\frac{M_0}{N};[\ell+r]_q\right),
\end{equation}
for $\Xi(\cdot):=(\xi(\cdot;l) )\in\Phi_{q,r}$. 
To proceed we need the Diophantine equation \eqref{eq_diof1} and the analogous one relating $q$ and $rN$ (recall from Remark~\ref{rk_101} that they are coprime), and stating the existance of  a unique pair of integers $\mu,\nu\in\Z$ such that 
\begin{equation}\label{eq_diof2}
\nu q+\mu(rN)=r,\qquad\qquad \textup{with} \qquad |\mu|<q .
\end{equation}
A simple computation yields 
\begin{equation}\label{eq_shift}
\frac{1}{q}\frac{M_0}{N}=\frac{M_0}{N}\left( \beta - \alpha\frac{r}{q}  \right)= \frac{M_0}{N}d_r + n_r \frac{M_0}{q}
\end{equation}
where
\begin{equation}\label{parametri}
d_r := \beta-\alpha\nu , \qquad \textup{and} \qquad  n_r:=- \mu\alpha r=\mu-\mu\beta q .
\end{equation}
Using this, the insertion of \eqref{eq_q014} in formula \eqref{eq_smart1} leads to
\begin{multline}\label{eq_smart3}
\langle\zeta^j_{(q,r)}(k); (V_{q,r}^{\nicefrac{M}{N}})^* \, \Xi\rangle  =
\\=\sqrt{\frac{|M_0|}{N}}\ \sum_{m\in\Z}\expo{\ii2\pi k_1m}\ 
\xi\left(\frac{M_0}{N}(k_2+j+d_r)+( m+n_r) \frac{M_0}{q};[mrN+r]_q\right) 
\\=\expo{- \ii2\pi k_1n_r}\ \sqrt{\frac{|M_0|}{N}}\ \sum_{m\in\Z}\expo{\ii2\pi k_1m}\ 
\xi\left(\frac{M_0}{N}(k_2+j+d_r)+ m \frac{M_0}{q};[mrN]_q\right) ,
\end{multline}
after a relabeling $m+n_r\mapsto m$ and using the fact that $[(m-n_r)rN+r]_q=[mrN]_q$ which follows from the Diophantine equations \eqref{eq_diof1} and \eqref{eq_diof2}.

The shift in the label $j$ due to the integer $d_r$ can be evaluated as in the Proposition~\ref{ppc} with $d_r=n_2$ there. Noting that now, we are considering the elements  $\zeta^j_{(q,r)}(k)$'s as basis vectors, this leads to 
\begin{equation}\label{matv}
V_{q,r}^{\nicefrac{M}{N}}(k) = \expo{\ii2\pi n_r k_1} \left( \num{V}_N(\expo{\ii2\pi q k_1} ) \right)^{{d_r}}, 
\end{equation}
with the matrix $\num{V}_N(\expo{\ii2\pi q k_1})$ defined in \eqref{eq_005bis}
(notice that here we get the transpose of the one in \eqref{eq_013}, as it should be).
One finds that $(V_{q,r}^{\nicefrac{M}{N}}(k))^N
=\expo{\ii2\pi k_1}\, \num{I}_N$, owing to $qd_r+n_rN=1$ using the parameters in \eqref{parametri}.
That the matrices \eqref{matu} and \eqref{matv} provides a representation of $\sss{A}_{\nicefrac{M}{N}}$ follows from the easy check (using the commutation relations \eqref{ecr}) that they commute up to the factor $\expo{\ii 2\pi\frac{M}{N}qd_r}$ with $qd_r=1-n_rN$ as before.

To proceed, we notice that a straightforward computation leads to
\begin{equation}\label{matvbis}
\qquad\qquad \num{L}_N(k)\ V_{q,r}^{\nicefrac{M}{N}}(k) \ \num{L}_N(k)^{-1} = \expo{\ii2\pi \frac{1}{N}k_1}\ 
(\num{V}_N)^{d_r}
\end{equation}
with the matrix $\num{V}_N$ in \eqref{shn}, and the conjugating $\num{L}_N(k)$ given by the diagonal matrix:
\begin{equation}\label{eq_matL}
 \num{L}_N(k):=\left(
\begin{array}{cccc}
1 & 0 &   \ldots  & 0 \\ 
0 & \expo{\ii 2\pi q k_1 \frac{1}{N} } &  \ldots  & 0 \\ 
\vdots &    & \ddots  & \vdots \\ 
0 & 0 &   \ldots &  \expo{\ii 2\pi q k_1 \frac{N-1}{N} }
                         \end{array}\right) .
\end{equation}
Clearly conjugation by $\num{L}_N(k)$ leaves the matrix $U_q(k)$ unchanged. 
Then irreducibility of the representation $\pi_k$ follows from the well know fact (cf. \cite[Lemma 1.8]{boca-01}) that, being $q$ and $N$ coprime, the $N \times N$ matrices $(\num{U}_N)^{qM}$ in \eqref{matu} and $(\num{V}_N)^{d_r}$ in \eqref{matvbis} generate the full matrix algebra $\text{Mat}_N(\C)$.

In fact, up to now, the variable $k$ needs not be restricted to the torus $\num{T}^2$.  However, the representations $\pi_k$ satisfy suitable pseudo-periodic conditions and they glue together to a \emph{bundle representation} $\widetilde\Pi_{q,r}$ of the torus algebra $\sss{A}_{\nicefrac{M}{N}}$, i.e. 
$$
\widetilde\Pi_{q,r}(\sss{A}_{\nicefrac{M}{N}})\subset\text{End}_{C(\num{T}^2)}(\Gamma(E_{N,q}))\simeq \Gamma(\text{End}(E_{N,q})) .
$$ Starting form the matrices  \eqref{matu}, and \eqref{matvbis} as well as \eqref{eq_013}, which can itself be written as 
$\num{G}_{N,q}(k) =  {}^t(\num{V}_N(\expo{\ii2\pi q k_1}) )$, by using the commutation relations \eqref{ecr} a simple computation shows that for any $k\in\R^2$ and $n:=(n_1,n_2)\in\Z^2$, it holds that
$$
U_q(k+n)=  \expo{\ii2\pi q \frac{M}{N}n_2}  U_q(k)=\num{G}_{N,q}(k)^{n_2}\ U_q(k)\ \num{G}_{N,q}(k)^{-n_2}
$$
and 
$$
V_{q,r}^{\nicefrac{M}{N}}(k+n)=V_{q,r}^{\nicefrac{M}{N}}(k)=
 \num{G}_{N,q}(k)^{n_2} \ V_{q,r}^{\nicefrac{M}{N}}(k) \ \num{G}_{N,q}(k)^{-n_2}  ,
$$
using here the commutativity of $\num{G}_{N,q}(k)$ with $V_{q,r}^{\nicefrac{M}{N}}(k)$. These \emph{pseudo-periodic}
conditions extend by linearity to the the whole algebra generated by $U_q(k)$ and $V_{q,r}^{\nicefrac{M}{N}}(k)$, namely
\begin{equation}\label{eq_p031}
\pi_{k+n}(a)=\num{G}_{N,q}(k)^{n_2}\ \pi_k(a)\ \num{G}_{N,q}(k)^{-n_2} 
\quad\quad \textup{with} 
\quad k\in\R^2 \ , \quad n=(n_1,n_2)\in\Z^2 .
\end{equation}
for any $a\in\sss{A}_{\nicefrac{M}{N}}$.
This condition assures that the map $A(\cdot):\R^2\to \bigsqcup_{k\in \num{R}^2}\text{End}_{\num C}(\sss{H}_{q,r}(k))$ given by $A(k):= \pi_{k}(a)$, for $a\in\sss{A}_{\nicefrac{M}{N}}$,  preserves the quasi periodicity of the frame $\boldsymbol{\zeta}_{(q,r)}(\cdot)$. That is, by writing $\tilde{\boldsymbol{\zeta}}_{(q,r)}(k):=A(k)\cdot \boldsymbol{\zeta}_{(q,r)}(k)$, in view of \eqref{eq_012} and \eqref{eq_p031}, it holds that 
\begin{multline*}
\tilde{\boldsymbol{\zeta}}_{(q,r)}(k+n)=A(k+n)\cdot \boldsymbol{\zeta}_{(q,r)}(k+n)=\num{G}_{N,q}(k)^{n_2}\cdot (A(k)\cdot \boldsymbol{\zeta}_{(q,r)}(k))=\num{G}_{N,q}(k)^{n_2}\cdot \tilde{\boldsymbol{\zeta}}_{(q,r)}(k) ,
\end{multline*}
for any $n:=(n_1,n_2)\in\Z^2$ and $k\in\R^2$. Thus,  the new frame $\tilde{\boldsymbol{\zeta}}_{(q,r)}(\cdot)$ describes the same vector bundle $E_{N,q}\to\num{T}^2$ as the starting one
$\boldsymbol{\zeta}_{(q,r)}(\cdot)$. Moreover, the map $A(\cdot)$ goes through the quotient $\num{T}^2$ and defines a map from $E_{N,q}$ to itself which preserves the fibers and commutes with the projection. That is $A(\cdot)\in\Gamma(\text{End}(E_{N,q}))$, thus providing a representation 
$\widetilde\Pi_{q,r}$ of the algebra $\sss{A}_{\nicefrac{M}{N}}$ as bundle endomorphisms. Indeed,  by its very definition $\widetilde\Pi_{q,r}(\cdot)$ is unitarily equivalent to the $(q,r)$-Weyl representation 
$\Pi_{q,r}$, i.e.  
$\widetilde\Pi_{q,r}(\cdot):= \bbb{F}_{q,r}\circ\Pi_{q,r}(\cdot) \circ{\bbb{F}_{q,r}}^{-1}$, and so it is faithful.

\section{Geometric duality and TKNN-equations}\label{sec_gem_dual}

The first step in proving the generalized TKNN-equations is to deduce a geometric duality between 
two different vector bundles associated to the same universal projection via two different bundle representations on the base space  $\num{T}^2$. The first of these is just the Bloch-Floquet representation $\widetilde\Pi_{q,r}(\cdot)$  discussed in the previous section while the second will be a naturally corresponding \virg{dual} one in a way we are about to describe. The geometric duality will relate suitable pullbacks of the vector bundles over the torus $\num{T}^2$.

\subsection{Geometric duality and untwisting functions}\label{sec_geoduality}

Let  ${E}_{N,q}\to\num{T}^2$ be again the rank $N$ Hermitian vector bundle generated by the frame of pseudo-periodic section $\boldsymbol{\zeta}_{(q,r)}(\cdot):=\{\zeta^j_{(q,r)}(\cdot)\}_{j=0,\ldots,N-1}$ as explained in Sect.~\ref{sec_loc_triv}. In the spirit of the Serre-Swan theorem, its fibers are given in \eqref{fib-can} via the projection $P_{N,q}(k)$ in \eqref{proj}. 
We also recall the fibered representation $\widetilde\Pi_{q,r}$ of $\sss{A}_{\nicefrac{M}{N}}$ to 
$\Gamma(\text{\upshape End}({E}_q))$ given in Theorem~\ref{teo_main_1}.

Now, to any projection $p\in\text{{\upshape Proj}}(\sss{A}_{\nicefrac{M}{N}})$ the projection-valued section $\tilde{\Pi}_{q,r}(p):=P(\cdot)$ associates a subbundle $ L_{q,r}(p)\subset E_{N,q}$ whose fibers, in parallel with \eqref{fib-can}, are given by
\begin{equation}\label{fib-can=qr}
\iota^{-1}(k) = P(k)\big(  \sss{H}_{q,r}(k) \big)=P(k) P_{N,q}(k) \big( \Phi_{q,r}^\ast \big). 
\end{equation}
This equation emphasizes the fact that  the twisting of the vector subbundle ${L}_{q,r}(p)$ is due to the twisting of the
\virg{environment} vector bundle ${E}_{N,q}$ coded by $P_{N,q}(\cdot)$, plus an extra twisting coming from the projection $P(\cdot)$.

Next, writing the projection $p=f(u,v)$ in $\sss{A}_{\nicefrac{M}{N}}$, for a suitable $f\in C^\infty(\num{T}^2)$, having in mind the form \eqref{matu} and \eqref{matv} of the representation, we look for an additional \virg{reference} bundle representation, $U_\text{ref}(\cdot)$ and $V_\text{ref}(\cdot)$, of $\sss{A}_{\nicefrac{M}{N}}$, for which
\begin{equation}\label{eq_proofGD4}
P(k_1,Nk_2)=f\left(U_q(Nk_2),V_{q,r}^{\nicefrac{M}{N}}(k_1)\right)=f\left(U_\text{ref}(M_0k_2),V_\text{ref}(k_1)\right)=P_\text{ref}(k_1,M_0k_2).
\end{equation}
Thus the projection-valued sections $P(\cdot)\in\Gamma(\text{End}({E}_{N,q}))$ and $P_\text{ref}(\cdot)\in\Gamma(\text{End}(\num{T}^2 \times \C^N))$  would be isomorphic realizations of the same universal projection $p$ in $\sss{A}_{\nicefrac{M}{N}}$

We are then lead to the operators
\begin{equation}\label{eq_ref_rep}
{U}_\text{ref}(k) :=\expo{\ii 2\pi k_2} \, (\num{U}_N)^{qM},\qquad\qquad 
{V}_\text{ref}(k):=V_{q,r}^{\nicefrac{M}{N}}(k)=\expo{\ii2\pi n_r k_1} \left( \num{V}_N(\expo{\ii2\pi q k_1} ) \right)^{{d_r}}, 
\end{equation}
where the numbers $d_r$ and $n_r$ are given in \eqref{parametri}.

Both ${U}_\text{ref}(\cdot)$ and ${V}_\text{ref}(\cdot)$ are elements in
$C(\num{T}^2)\otimes\text{\upshape Mat}_N(\C)\simeq C(\num{T}^2;\text{\upshape Mat}_N(\C))$. 
The map $\Pi_\text{ref}(u)={U}_\text{ref}(\cdot)$, and $\Pi_\text{ref}(v)={V}_\text{ref}(\cdot)$ defines a $\ast$-isomorphism between $\sss{A}_{\nicefrac{M}{N}}$ and $C^\ast({U}_\text{ref}(\cdot),{V}_\text{ref}(\cdot))\simeq C(\num{T}^2;\text{\upshape Mat}_N(\C))$, that we name {(for lack of a better name)} the \emph{reference bundle representation} of the NCT-algebra $\sss{A}_{\nicefrac{M}{N}}$. Indeed, one easily checks that ${U}_\text{ref}(\cdot){V}_\text{ref}(\cdot)=\expo{\ii2\pi\frac{M}{N}}{V}_\text{ref}(\cdot){U}_\text{ref}(\cdot)$. 
Then surjectivity of $\Pi_\text{ref}$ follows from the universal property of 
$\sss{A}_{\nicefrac{M}{N}}$. Observing that $({U}_\text{ref}(k))^N=  
\expo{\ii 2\pi k_2 \,N}
\otimes\num{I}_N$, and $({V}_\text{ref}(k))^N= \expo{\ii 2\pi k_1} \otimes\num{I}_N$ and so $C^\ast({U}_\text{ref}(\cdot)^N,{V}_\text{ref}(\cdot)^N)\simeq C(\num{T}^2)$, injectivity then follows from
\cite[Prop.~1.11]{boca-01}.

We view the operators ${U}_\text{ref}(\cdot)$ and ${V}_\text{ref}(\cdot)$ (and all elements in $C^\ast({U}_\text{ref}(\cdot),{V}_\text{ref}(\cdot))$ for what it matters) as acting on the trivial bundle $\num{T}^2 \times \C^N$ over $\num{T}^2$. Then, any $p\in\text{{\upshape Proj}}(\sss{A}_{\nicefrac{M}{N}})$ is mapped by $\Pi_\text{ref}$ to a  projection-valued section $\Pi_\text{ref}(p):=P_\text{ref}(\cdot)$ which defines a vector subbundle ${L}_\text{ref}(p)\to\num{T}^2$ of the trivial vector bundle $\num{T}^2 \times \C^N$. 

Then, equation \eqref{eq_proofGD4} leads to compare suitable pullback of vector bundles , namely
\begin{align}\label{eq_proofGD5}
\varphi^\ast_{(1,M_0)}{L}_\text{ref}({p}) &\simeq\ \bigsqcup_{k\in\num{T}^2}\ P_\text{ref}(k_1,M_0k_2)\, \C^N , \nonumber \\
\varphi^\ast_{(1,N)}{L}_{q,r}({p}) &\simeq\ \bigsqcup_{k\in\num{T}^2}\ P(k_1,Nk_2) P_{N,q}(k_1,Nk_2)  \big( \Phi_{q,r}^\ast \big) .  
\end{align}
From \eqref{eq_proofGD5} it follows that, for any $k\in\num{T}^2$, the two corresponding  fibers 
of $\varphi^\ast_{(1,M_0)}{L}_\text{ref}({p})$ and $\varphi^\ast_{(1,N)}{L}_{q,r}({p})$
are determined by the same projection \eqref{eq_proofGD4} acting on a $N$-dimensional complex vector space. However for $\varphi^\ast_{(1,M_0)}{L}_\text{ref}({p})$
such a  vector space does not depend on $z$ and coincides with $\C^N$. Conversely, for  $\varphi^\ast_{(1,N)}{L}_{q,r}({p})$ the vector space in which the projection acts depends on $z$  as a consequence of the non triviality of the \virg{environment} vector bundle ${E}_{N,q}$. In other words 
$\varphi^\ast_{(1,M_0)}{L}_\text{ref}({p})$ and $\varphi^\ast_{(1,N)}{L}_{q,r}({p})$ coincide locally but  the latter vector bundle has an extra twist induced by the rank $N$ projection 
$(P_{N,q}\circ \varphi_{(1,N)})(\cdot)$. 

\goodbreak
\medskip

The projection $(P_{N,q}\circ \varphi_{(1,N)})(\cdot)$ gives the pullback vector bundle
$\varphi_{(1,N)}^\ast{E}_{N,q}\to\num{T}^2$ the structure of which 
can be determined by means of the pullback of the frame  $\boldsymbol{\zeta}_{(q,r)}(\cdot)$. Indeed, 
from the very definition of pullback, the fiber space $(\varphi_{(1,N)}^\ast{E}_{N,q})_k$ over the point $k$ coincides with the fiber space $({E}_{N,q})_{\varphi_{(1,N)}(k)}$ over the transformed point 
$\varphi_{(1,N)}(k)$ and the latter is spanned by the family of sections $\varphi_{(1,N)}^\ast\boldsymbol{\zeta}_{(q,r)}:=\boldsymbol{\zeta}_{(q,r)}\circ \varphi_{(1,N)}$ evaluated at the point $k$. 
By means of \eqref{eq_012}, a simple computation shows that for any $k\in\R^2$ and $n\in\Z^2$ one has 
\begin{equation}\label{eq_GD_01}
(\varphi_{(1,N)}^\ast\boldsymbol{\zeta}_{(q,r)})(k+n)=\boldsymbol{\zeta}_{(q,r)}(k_1+n_1 ,N(k_2+n_2))=(\num{G}_{N,q}(k))^{Nn_2}\cdot\varphi_{(1,N)}^\ast\boldsymbol{\zeta}_{(q,r)}(k). 
\end{equation}
This shows that the frame $\varphi_{(1,N)}^\ast\boldsymbol{\zeta}_{(q,r)}$ is pseudo-periodic for the action of the matrix
\begin{equation}\label{eq_GD_02}
\num{G}_{N,q}(k)^{N}= \expo{\ii 2\pi qk_1}\ \num{I}_N= {(-1)}^{N-1}\ \text{det}\left[\num{G}_{N,q}(k)\right] \num{I}_N.
\end{equation}
Equation \eqref{eq_GD_02} means that any single section of the frame $\varphi_{(1,N)}^\ast{\zeta}_{(q,r)}^j$, with $j=0,\ldots,N-1$,
is pseudo-periodic with respect to the phase factor ${(-1)}^{N-1} \text{det}\left[\num{G}_{N,q}(k)\right]$. Therefore, 
any section $\varphi_{(1,N)}^\ast{\zeta}_{(q,r)}^j$ defines a line bundle (rank 1 vector bundle) over 
$\num{T}^2$. The transition functions
$\boldsymbol{\ell}:=\{\ell_{a,b}\}_{a,b=1,\ldots,4}$ subordinate to the minimal cover \eqref{RA43''} of $\num{T}^2$ are then given by $\ell_{a,b}(k):={(-1)}^{N-1} \text{det}\left[g_{a,b}(k)\right]$
where ${\boldsymbol{g}}:=\{{g}_{a,b}\}_{a,b=1,\ldots,4}$ are the transition functions of the bundle $E_{N,q}$  according to \eqref{RA40} and \eqref{RA41}. The corresponding line bundle is the \emph{determinant line bundle} $\text{det}({E}_{N,q})\to \num{T}^2$.
The sign ${(-1)}^{N-1}$ does not enter in the transformation formula for connections such as in \eqref{eq_con1} and so it does not  affect the computation of any curvature and corresponding Chern numbers. 

Summarizing, one  obtains the following relation:
\begin{equation}\label{eq_GD_03}
\varphi_{(1,N)}^\ast{E}_{N,q} \simeq\ \underbrace{\text{det}({E}_{N,q})\oplus\ldots\oplus\text{det}({E}_{N,q})}_{N-\text{times}}\ \simeq\ \left(\text{det}({E}_{N,q}) \right)^{\oplus N}.
\end{equation}

\Proof {\bf of Theorem \ref{teo_main_3}.}
As mentioned, \eqref{eq_proofGD5} yields that the bundles
$\varphi^\ast_{(1,M_0)}{L}_\text{ref}({p})$ and $\varphi^\ast_{(1,N)}{L}_{q,r}({p})$ coincide locally but  the latter vector bundle has an extra twist induced by the vector bundle 
$\varphi^\ast_{(1,N)}{E}_{N,q}\simeq \left(\text{det}({E}_{N,q}) \right)^{\oplus N}$. Then, if $Q_{N,q}(\cdot)$ is the rank one projection defining the line bundle $\text{det}(E_{N,q})$, the identification \eqref{eq_GD_03} leads to $(P_{N,q}\circ \varphi_{(1,N)})(\cdot)=\num{I}_N\otimes Q_{N,q}(\cdot)$ where $\num{I}_N$ denotes the rank $N$ constant projection. 
It follows that
\begin{align*}
(P\ P_{N,q}\circ \varphi_{(1,N)})(\cdot)&=[(P\circ \varphi_{(1,N)})(\cdot)\otimes\num{I}_{1}]\ [\num{I}_N\otimes Q_{N,q}(\cdot)]=(P_{\text{ref}}\circ \varphi_{(1,M_0)})(\cdot)\otimes Q_{N,q}(\cdot).
\end{align*}
For the corresponding vector bundles, this is just the isomorphism \eqref{eq_GD_04}:
$$
 \varphi_{(1,N)}^\ast L_{q,r}(p) \simeq\  \varphi_{(1,M_0)}^\ast{L}_\text{{\upshape ref}}(p)\otimes\text{\upshape det}({E}_{N,q}).
$$
\CVD


\subsection{The reference bundle representation}\label{sec_RBR}
We are finally ready to prove Theorem~\ref{teo:new1}. To this end, we just need the following.
 \begin{propos}\label{prop_abst_chern}
 For any $p\in\text{{\upshape Proj}}(\sss{A}_{\nicefrac{M}{N}})$ the associated vector bundle 
 ${L}_\text{{\upshape ref}}(p)\to\num{T}^2$ has rank $\text{\upshape Rk}({L}_\text{{\upshape ref}}(p)):=N \ncint({p})$ and first Chern number $C_1({L}_\text{{\upshape ref}}(p))=\ \ncC({p})$
where $\ncC$\, is defined by equation \eqref{eq_009}.
\end{propos}
\Proof 
In fact, we compute rank and first Chern number of a simpler bundle. With the matrix $\num{L}_N$ in 
\eqref{eq_matL}, out of \eqref{eq_ref_rep},  for any $k=(k_1,k_2)\in\R^2$ we get  
\begin{align}\label{eq_ref_rep3}
{\widetilde{U}}_\text{ref}(k_1,k_2) &:= \num{L}_N(qNk_1)\ {U}_\text{ref}(k_2)\ \num{L}_N(qNk_1)^{-1}  = 
\expo{\ii 2\pi k_2} \, (\num{U}_N)^{qM}  
\nonumber \\
{\widetilde{V}}_\text{ref}(k_1,k_2) &:= \num{L}_N(qNk_1)\ {V}_\text{ref}(Nk_1)\ \num{L}_N(qNk_1)^{-1}  =
\expo{\ii2\pi k_1} \left( \num{V}_N \right )^{d_r} 
\end{align}
Clearly, having conjugated by a unitary matrix, we get as before a 
$\ast$-isomorphism $\widetilde{\Pi}_\text{ref}$ between $\sss{A}_{\nicefrac{M}{N}}$
and $C^\ast({\widetilde{U}}_\text{ref}(\cdot),{\widetilde{V}}_\text{ref}(\cdot))\simeq C(\num{T}^2;\text{\upshape Mat}_N(\C))$ given by
$\Pi_\text{ref}(u)={\widetilde{U}}_\text{ref}(\cdot)$ and 
$\Pi_\text{ref}(v)={\widetilde{V}}_\text{ref}(\cdot)$ on generators.
But with the simpler form before, we have in addition a natural identification \cite[Cor.~1.12]{boca-01}:
\begin{equation}\label{id-int}
\nint \ = \int_{\num{T}^2} \ \dd z\ \left( \frac{1}{N} \text{Tr}_N \right) \circ \widetilde{\Pi}_\text{ref} ,
\end{equation}
where $\dd z = \dd k_1\wedge \dd k_2$ is the Haar measure on $\num{T}^2$. 
This is none other that the computation
$$
\frac{1}{N}\int_{\num{T}^2} \ \dd k_1\wedge \dd k_2 \ 
\text{Tr}_N\left({\widetilde{U}}_\text{ref}(k)^n{\widetilde{V}}_\text{ref}(k)^m\right)
=\delta_{n,0}\ \delta_{m,0}=\nint({u}^n{v}^m) .
$$ 
Finally, using the definition \eqref{eq_008} of the derivations, one checks that
\begin{equation}\label{id-der}
\widetilde{\Pi}_\text{ref}\circ \ncpartial_1=(\partial_{k_2} )\circ \widetilde{\Pi}_\text{ref} \qquad \textup{and}\qquad 
\widetilde{\Pi}_\text{ref}\circ \ncpartial_2=(\partial_{k_1} )\circ\widetilde{\Pi}_\text{ref} .
\end{equation}

Next, if $p\in \text{{\upshape Proj}}(\sss{A}_{\nicefrac{M}{N}})$ with corresponding projection 
$\widetilde{P}_\text{ref}(\cdot):=\widetilde{\Pi}_\text{ref}(p)$ from equations \eqref{eq_ref_rep3} one deduces that
$\widetilde{P}_\text{ref}(k_1,k_2)=\num{L}_N(qNk_1)\ P_\text{ref}(Nk_1,k_2)\ \num{L}_N(qNk_1)^{-1}$. Since the unitary matrix $\num{L}_N(qNk_1)$ is periodic in $k_1$, it defines a globally trivial change of the orthonormal frame on the fibers of the vector bundle determined by $P_\text{ref}(\cdot)$. In other words
the vector bundles determined by the projections $\widetilde{P}_\text{ref}(\cdot)$ and $P_\text{ref}(\cdot)$ are related has 
\begin{equation}\label{eq_ref_rep4}
 {\widetilde{L}}_\text{ref}(p)\simeq \varphi_{(N,1)}^\ast{L}_\text{ref}(p) ,
\end{equation}
where $\varphi_{(N,1)}$ is the  continuous map $\varphi_{(N,1)}:\num{T}^2\to \num{T}^2$ defined by \eqref{eq_appBBB}.

It is shown in \cite[Cor.~1.22]{boca-01} that the function $\num{T}^2 \ni k \mapsto 
\text{Tr}_N(\widetilde{P}_\text{ref}(k))$ is locally constant (and in fact constant). This is just the rank of the bundle  ${\widetilde{L}}_\text{{\upshape ref}}(p)$. Using the identification \eqref{id-int}, this leads to
$$
\text{Rk}(\widetilde{L}_\text{{\upshape ref}}(p)) := \text{Tr}_N(\widetilde{P}_\text{ref}) = N \nint({p}) .
$$
In turn, since the pullback preserves the rank, from \eqref{eq_ref_rep4} one gets 
$\text{Rk}(L_\text{{\upshape ref}}(p)) = N \ncint({p})$. 
 
Finally, using the identifications \eqref{id-der} and \eqref{id-int} and the definition \eqref{eq_009} 
for $\ncC({p})$,  the first first Chern number 
$C_1({\widetilde{L}}_\text{{\upshape ref}}(p))$ of the bundle 
${\widetilde{L}}_\text{{\upshape ref}}(p)$  is computed to be 
\begin{align*}
C_1({\widetilde{L}}_\text{{\upshape ref}}(p)) &:= \frac{\ii}{2\pi }\int_{\num{T}^2}  
\text{Tr}_N \left( \widetilde{P}_\text{ref} (\dd \widetilde{P}_\text{ref})^2 \right) \\
&\ =-\frac{\ii }{ 2\pi}\nint p \big(\ncpartial_1(p)
\ncpartial_2(p)-\ncpartial_2(p)\ncpartial_1(p) \big) = N \ncC(p).
\end{align*}
In turn, from the equivalence \eqref{eq_ref_rep4}, using \eqref{eq_appBBB1}, one gets 
$C_1({L}_\text{ref}(p))=\ \ncC({p})$
\CVD

\goodbreak
\bigskip

\appendix

\section{The computation of the Chern number}\label{se:ccn}

The frame $\boldsymbol{\zeta}_{(q,r)}(\cdot)$ with the pseudo-periodic conditions \eqref{eq_012} encode all topological properties of the vector bundle $\iota: {E}_{N,q}\to\num{T}^2$.
Indeed, the unitary matrix $\num{G}_{N,q}(k)$ in \eqref{eq_013} yields the transition functions of the 
bundle $E_{N,q}$.
The base manifold being $\num{T}^2 = \num{S}^1 \times \num{S}^1$ it has a minimal cover made by four open sets, 
$\{O_a\}_{a=1,\ldots,4}$, each diffeomorphic to a open square in $\R^2$, that can be taken to be 
\begin{align}\label{RA43''} 
&O_1=\{z(k)\in\num{T}^2 \ : \ -\epsilon<k_1<\nicefrac{1}{2}+\epsilon,\ -\epsilon<k_2<\nicefrac{1}{2}+\epsilon\}\nonumber\\
&O_2=\{z(k)\in\num{T}^2 \ : \ \nicefrac{1}{2}-\epsilon<k_1<1+\epsilon,\ -\epsilon<k_2<\nicefrac{1}{2}+\epsilon\}\\
&O_3=\{z(k)\in\num{T}^2 \ : \ -\epsilon<k_1<\nicefrac{1}{2}+\epsilon,\ \pi-\epsilon<k_2<1+\epsilon\}\nonumber\\
&O_4=\{z(k)\in\num{T}^2 \ : \ \nicefrac{1}{2}-\epsilon<k_1<1+\epsilon,\ \nicefrac{1}{2}-\epsilon<k_2<1+\epsilon\} , \nonumber
\end{align}
with $z(k):=(\expo{\ii2\pi k_1},\expo{\ii2\pi k_2})$ and a small enough $\epsilon>0$.  Any intersection $O_{a,b}:=O_a\cap O_b$ is non empty, and is made by the union of two disjoint sets. 

By restricting the frame $\boldsymbol{\zeta}_{(q,r)}(\cdot)$ to each open set $O_a$ we get diffeomorphisms 
$$ 
\varphi_{a} :  {O_a}\times\C^N \longrightarrow \iota^{-1} (O_a)\subset E_{N,q} ,
$$ 
defined by
\begin{equation}\label{appC_eq1}
\varphi_{a}(k,\text{v}) = {\zeta_{(q,r)}^0}|_{O_a}(k)\ v_0 +\ldots+ {\zeta_{(a)}^{N-1}}|_{O_a}(k)\ v_{N-1}\ ,
\end{equation}
where $\text{v}=(v_0,\ldots,v_{N-1})$, with corresponding transition functions $g_{a,b}(k):=\left.\varphi_{a}^{-1}\circ\varphi_{b}\right|_k$ on the intersections $g_{a,b}: O_{a,b}=O_a\cap O_b\to\text{U}(N)$, the latter being the group of unitary $N\times N$ matrices. 
For each $k$ these obey the usual conditions:
\begin{equation}\label{eq_tfpro}
g_{a,a}(k)=\num{I}_N, \qquad g_{b,a}(k) = (g_{a,b}(k))^{-1}  \qquad \textup{and} 
\qquad\ g_{a,b}(k)\cdot g_{b,c}(k)=g_{c,a}(k).
\end{equation}
By using the pseudo-periodic conditions in \eqref{eq_012}, a straightforward computation yields the following for the 
transitions functions of the vector bundle. First of all,
\begin{equation}\label{RA40}
 g_{1,2}(k)=\num{I}_N=g_{3,4}(k) , 
\end{equation}
which means that  the vector bundle is trivial in the direction $k_1$. On the other hand, 
\begin{equation}\label{RA41}
g_{a,b}(k)=\left\{
\begin{aligned}
&\num{I}_N&&\text{if}\ \ k\in O_{a,b}(k_2\sim\nicefrac{1}{2})\\
&{^t\num{G}_{N,q}}(k)&&\text{if}\ \ k\in O_{a,b}(k_2\sim0)
\end{aligned}
\right.\ \ \ \ \ \ \ \ \text{with}\ \ (a,b)=(1,3),\ (1,4),\ (2,3),\ (2,4)
\end{equation}
which means that  the vector bundle is twisted by $\num{G}_{N,q}$ in the direction $k_2$.
In the above, $O_{a,b}=O_{a,b}(k_2\sim\nicefrac{1}{2})\cup O_{a,b}(k_2\sim 0)$, where $O_{a,b}(k_2\sim\nicefrac{1}{2})$ is a strip around  $k_2=\nicefrac{1}{2}$ and $O_{a,b}(k_2\sim 0)$ is a strip around $k_1=0$, both $2\epsilon$ wide and defined modulo $\Z$.

\goodbreak
\medskip
To compute the first Chern number of the  vector bundle $E_{N,q}$ one may use a connection on the bundle, the result non depending on the particular connection. Any such a connection can be given as a collection $\omega=\{\omega_a\}$, of 1-forms on  the open sets $O_a$, for $a=1,\ldots,4$, with values in the Lie algebra $\rrr{u}(N)$ of anti-Hermitian $N\times N$ matrices (the Lie algebra of the structure group $\text{U}(N)$), glued together by the transition functions, i.e.
\begin{equation}\label{eq_con1}
\omega_{a}=g_{a,b}\ \dd{g_{a,b}}^{-1}+{g_{a,b}}\ \omega_{b} \ {g_{a,b}}^{-1},\qquad\qquad a,b=1,\ldots,4 .
\end{equation}
Using the specific form of the transition functions \eqref{RA40} and \eqref{RA41}, the consistency equation \eqref{eq_con1} can be rewritten in terms of pseudo-periodic conditions as
\begin{align}\label{eq_con2}
\omega(k_1+1,k_2) &= \omega(k_1,k_2) , \nonumber \\
\omega(k_1,k_2+1) &= {\overline{\num{G}}_{N,q}}(k_1)\ \omega(k_1,k_2)\ {^t\num{G}_{N,q}}(k_1) + 
{\overline{\num{G}}_{N,q}}(k_1)\ \dd {^t\num{G}_{N,q}}(k_1) .
\end{align}
By its very definition, the Berry's connection $\omega^{(\text{B})}$ in \eqref{eq_berry_con} verifies the consistency rule \eqref{eq_con2} as can be verified. Also,
a direct computation shows that the matrix valued 1-form
\begin{equation}\label{ch5_harp_connection}
\omega^{(N,q)}(k_1,k_2):=\ii\frac{2\pi q}{ N}\left(
{k_2} \num{I}_N + A \right)\dd k_1, \qquad 
A= \left( \begin{array}{cccc}
0 & 0 &  \ldots  & 0 \\ 
0 & 1 & \ldots& 0 \\ 
\vdots & \vdots &  \ddots & \vdots  \\ 
0 & 0 &  \ldots  &  (N-1)
                         \end{array}\right)
\end{equation}
verifies
  \eqref{eq_con2} as well.  
 Since $\omega^{(N,q)}$ has only component $\dd k_1$ one has $\omega^{(N,q)}\wedge\omega^{(N,q)}=0$. The corresponding curvature $K^{(N,q)}:=\dd (\omega^{(N,q)})$ is the globally defined (and constant) $\rrr{u}(N)$-valued $2$-form given by
\begin{equation}\label{RAharcurv}
K^{(N,q)}(k_1,k_2)=\left(\dfrac{2\pi q}{\ii N}\ \num{I}_N\right)\ \dd k_1\wedge \dd k_2 .
\end{equation}
Upon integrating we get 
$$
C_1(E_{N,q})= \frac{\ii}{2\pi} \int_{\num{T}^2} \text{Tr}_N[K^{(N,q)}] = q  
$$
as the Chern number of the vector bundle $E_{N,q}$.


\end{document}